%% file: main.tex
\documentclass{article}

\usepackage[main, final]{neurips_2026}

\input{inc/inc_packages}
\input{inc/inc_macros}

\addtocontents{toc}{\setcounter{tocdepth}{-10}}

\title{\beastthree: Animal behavioral analysis and neural encoding from multi-view video via Gaussian splatting}

\author{%
  Yanchen Wang\\
  Columbia University\\
  \texttt{yw4503@columbia.edu}
  \And
  Lenny Aharon\\
  Columbia University\\
  \texttt{lenny.aharon@columbia.edu}
  \And
  Wangshu Zhu\\
  Columbia University\\
  \texttt{wz2708@columbia.edu}
  \AND
  Kyle Daruwalla\\
  Cold Spring Harbor\\
  \texttt{daruwal@cshl.edu}
  \And
  Linghua Zhang\\
  Cold Spring Harbor\\
  \texttt{lingzhan@cshl.edu}
  \And
  Jiaru Zou\\
  Stanford University\\
  \texttt{jiaru@stanford.edu}
  \And
  Selmaan Chettih\\
  Columbia University\\
  \texttt{sc4551@columbia.edu}
  \And
  Helen Hou\\
  Cold Spring Harbor\\
  \texttt{hou@cshl.edu}
  \AND
  Liam Paninski\\
  Columbia University\\
  \texttt{lmp2107@columbia.edu}
  \And
  Matthew R Whiteway\\
  Columbia University\\
  \texttt{m.whiteway@columbia.edu}
}

\begin{document}

\maketitle

\input{contents/c0_abstract}

\input{contents/c1_introduction}

\input{contents/c2_related_work}

\input{contents/c3_methods}

\input{contents/c4_experiments}

\input{contents/c5_conclusion}

\clearpage
\newpage
\input{contents/c6_acks}

\bibliographystyle{unsrt}
\bibliography{references}


\clearpage
\newpage
\addtocontents{toc}{\setcounter{tocdepth}{1}}
\appendix
\input{appendix/main}


\newpage
\input{checklist}

\end{document}

%% file: inc/inc_packages.tex
\usepackage[utf8]{inputenc}     
\usepackage[T1]{fontenc}        
\usepackage{hyperref}           
\usepackage{url}                
\usepackage{booktabs}           
\usepackage{adjustbox}          
\usepackage{amsfonts}           
\usepackage{amsmath}            
\usepackage{nicefrac}           
\usepackage{microtype}          
\usepackage{xcolor}             
\usepackage{graphicx}           
\usepackage{subcaption}         
\usepackage{wrapfig}            
\usepackage{algorithm}          
\usepackage{algorithmic}        
\usepackage{multirow}           
\usepackage{soul}               
\usepackage{xspace}             
\usepackage{caption}            
\hypersetup{
    colorlinks,
    linkcolor={blue!64!black},
    citecolor={blue!64!black},
    urlcolor={blue!64!black}
}

\usepackage{enumitem}

%% file: inc/inc_macros.tex
\newcommand{\beast}{\textsc{Beast}\xspace}
\newcommand{\beastthree}{\textsc{Beast3D}\xspace}

\definecolor{MyDarkBlue}{rgb}{0,0.08,0.8}

%% file: contents/c0_abstract.tex
\begin{abstract}
Multi-view video recordings are increasingly used to capture the 3D movements of animals in experimental settings, yet extracting rich 3D representations from these recordings remains challenging. 
Supervised pose estimation requires extensive manual annotation, while general-purpose 3D reconstruction models trained on generic scene datasets fail on the specialized imagery and sparse-view setting of laboratory experiments. 
We address these limitations with \beastthree, a self-supervised pretraining framework that learns 3D visual representations from unlabeled, calibrated multi-view video. 
\beastthree uses a vision transformer to predict 3D Gaussian splats that reconstruct held-out views through differentiable rendering, while simultaneously segmenting the animal from the background.
\beastthree reconstructs 3D structure with as few as four views by conditioning directly on known camera parameters---unlike general-purpose models, which must estimate camera geometry from dense overlapping viewpoints that are seldom available in lab settings.
Through comprehensive evaluation across four species, we demonstrate that \beastthree produces rich, viewpoint-invariant features that transfer effectively to three downstream tasks: 
\textit{novel view synthesis}, which validates the quality of the learned 3D representations; 
\textit{multi-view pose estimation}, which provides the sparse keypoint trajectories widely used in behavioral analysis;
and \textit{neural encoding}, which relates 3D behavioral features to simultaneously recorded neural activity.
\beastthree thus establishes a versatile framework for behavioral analysis that leverages 3D structure in modern multi-view laboratory recordings. Our project page and code: \url{https://ppwangyc.github.io/projects/beast3d}.
\end{abstract}

%% file: contents/c1_introduction.tex
\section{Introduction}
\label{sec:intro}

Advances in neuroscience increasingly depend on precise quantification of animal movement, yet the single-view video recordings that dominate current practice are fundamentally limited: self-occlusions obscure body parts, and 2D observations cannot recover the full 3D kinematics of movement~\citep{marshall2022leaving}. A growing number of laboratories address these limitations by recording with multiple synchronized and calibrated cameras, enabling accurate 3D measurements of constrained and freely moving animals~\citep{ marshall2021continuous, chettih2024barcoding, haakansson2024application, klibaite2025mapping, ulutas2025high}. 
The resulting multi-view data support a range of 3D analysis tasks, from sparse keypoint estimation that describes posture and joint kinematics~\citep{gunel2019deepfly3d, bala2020automated, dunn2021geometric, karashchuk2021anipose, monsees2022estimation, han2024multi, cheng2025real, aharon2026lightning} to dense 3D meshes that capture the full surface geometry and articulation of the subject~\citep{zuffi20173d, badger20203d, bohnslav2023armo}. 
These explicit 3D representations offer a natural basis for relating behavior to neural activity, providing an interpretable and low-dimensional description of animal movement independent of camera viewpoint.

\begin{figure*}[t!]
\begin{center}
  \includegraphics[width=\linewidth]{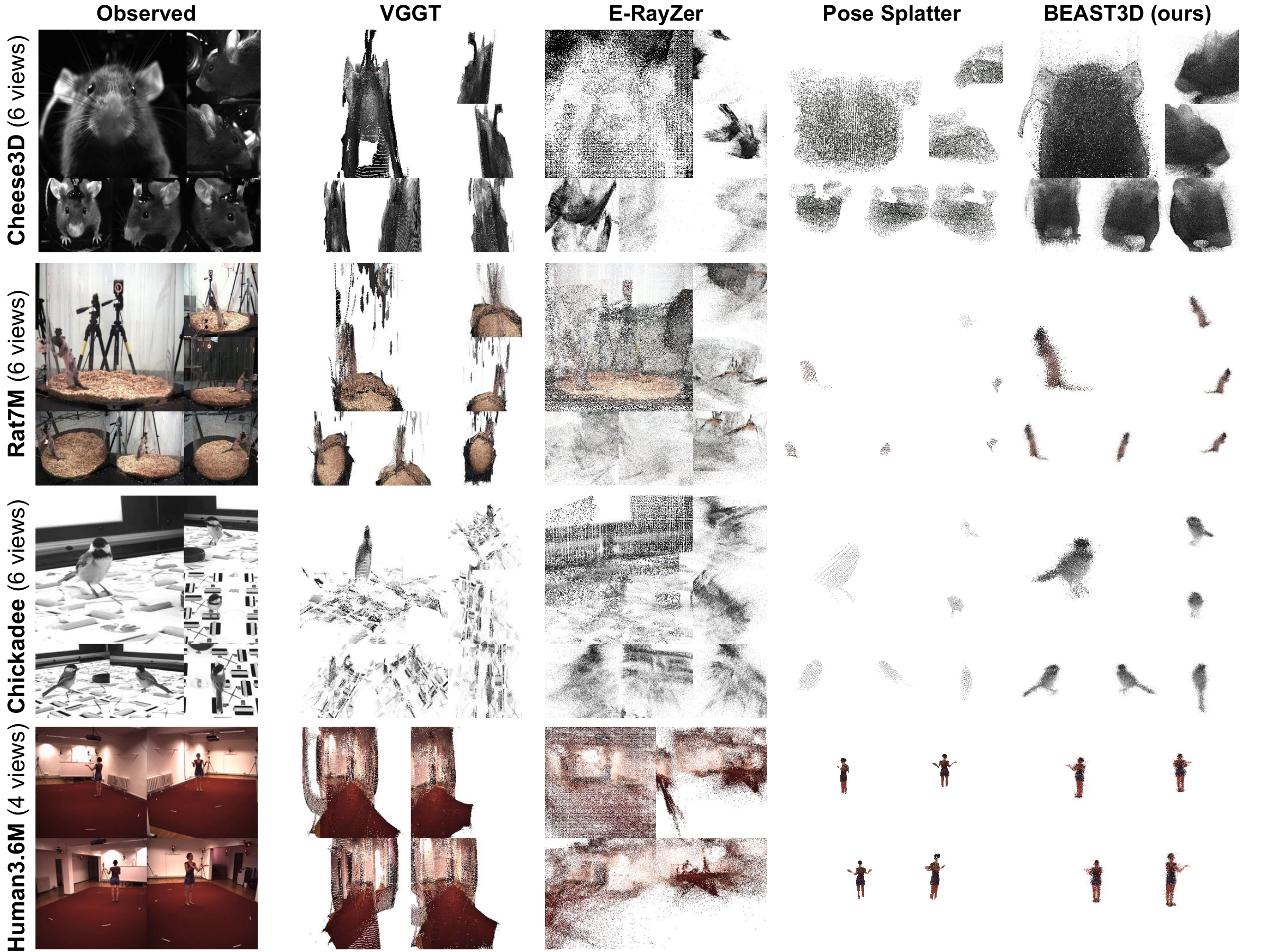}
\end{center} 
\caption{\textbf{3D point clouds from \beastthree and leading baselines.} An example scene from diverse datasets (\textit{left} column; Cheese3D~\citep{daruwalla2026cheese3d}, Rat7M~\citep{marshall2021continuous}, Chickadee~\citep{chettih2024barcoding}, Human3.6M~\citep{ionescu2013human3}) is encoded into a 3D point cloud by general-purpose models (VGGT~\citep{wang2025vggt}, E-RayZer~\citep{zhao2025erayzer}) and tailored per-dataset models (Pose Splatter~\citep{goffinet2025pose}, \beastthree). \beastthree achieves strong performance while simultaneously providing foreground segmentation of the subject. The point cloud positions are predicted by each model, and each point is colored by the corresponding color on each pixel.
}
\label{fig:fig1_baselines_compare}
\end{figure*}

Despite the richness of multi-view recordings, current approaches to extracting 3D representations face important limitations. Supervised pose estimation methods require extensive manual annotation of keypoints, which is labor-intensive and must be repeated for each new experimental setup or species. 
Mesh-based approaches such as SMAL~\citep{zuffi20173d} provide richer surface representations, but require a species-specific template mesh and costly per-frame optimization, limiting their scalability across species and long recordings~\citep{badger20203d, li2021coarse} (though see~\citep{hu2026sam}).
Meanwhile, recent 3D computer vision models such as VGGT~\citep{wang2025vggt} and E-RayZer~\citep{zhao2025erayzer} can produce dense 3D reconstructions from images in a single forward pass, but these methods are trained on generic scene datasets and perform poorly on the specialized imagery of laboratory recordings: close-up views of animals under controlled lighting from fixed camera rigs (Fig.~\ref{fig:fig1_baselines_compare}). Furthermore, these models are designed for settings with dense, highly overlapping viewpoints, and devote substantial capacity to estimating unknown camera parameters, which can fail with limited view overlap (Fig.~\ref{fig:camera_pose}). Neither assumption holds in typical laboratory camera rigs, which use a small number of widely spaced cameras with known calibration.

We address these gaps with \beastthree, a self-supervised pretraining framework that learns 3D visual features from calibrated multi-view animal behavior videos. Given only unlabeled, synchronized multi-view images and corresponding camera parameters, \beastthree predicts a set of 3D Gaussian splats via a transformer trained to reconstruct held-out views. 
The model simultaneously segments the animal from the background by distilling masks from a video segmentation model, producing clean foreground representations. 
The self-supervised pretraining produces a flexible backbone that can be fine-tuned for multiple analytical needs.
We demonstrate the quality of \beastthree's features by comparing against state-of-the-art models across four species (mouse, rat, chickadee and human), showing substantially improved reconstructions with as few as four views (Fig.~\ref{fig:fig1_baselines_compare}). 
We evaluate these features on three downstream tasks: 
(i) \textit{novel view synthesis}, where a 3D scene constructed from a subset of views must reconstruct held-out views; 
(ii) \textit{multi-view pose estimation}, where a lightweight head trained on \beastthree features predicts a sparse set of keypoints;
and (iii) \textit{neural encoding}, where \beastthree's features predict neural activity in the mouse and chickadee datasets. 
In all cases \beastthree achieves competitive or superior performance compared to baselines including VGGT and DINOv3~\citep{simeoni2025dinov3}. 
These results establish \beastthree as a versatile framework for extracting rich 3D features from multi-view laboratory recordings, bridging the gap between general-purpose 3D computer vision and the specialized demands of behavioral neuroscience.

%% file: contents/c2_related_work.tex
\section{Related Work}
\label{sec:related}

\textbf{Self-supervised visual representation learning.}
\label{sec:related:ssl}
Self-supervised pretraining has become the dominant paradigm for learning transferable visual features.
Masked image modeling approaches, exemplified by MAE~\citep{he2022masked}, train a vision transformer~\citep{dosovitskiy2020image} to reconstruct randomly masked pixel patches.
An alternative family of methods avoids pixel-level reconstruction.
I-JEPA~\citep{assran2023self} predicts latent representations of masked regions from context, avoiding the low-level pixel bias of MAE-style objectives.
The DINO family~\citep{caron2021emerging,oquab2023dinov2, simeoni2025dinov3} learns representations through self-distillation with momentum teachers, producing features that capture semantic structure without explicit reconstruction.
These methods learn powerful 2D features but 
have no mechanism for reasoning about 3D structure.

\textbf{3D scene modeling.}
\label{sec:related:3dgs}
3D Gaussian Splatting (3DGS; ~\cite{kerbl20233d}) 
represents a scene as a set of 3D Gaussians---each with its own position, shape and color parameters---rendered via differentiable rasterization, achieving high visual fidelity. 
A key limitation of 3DGS (as well as earlier neural radiance field approaches,~\cite{mildenhall2021nerf}) is the need for per-scene optimization: each new scene requires minutes of gradient descent to fit the representation.
Recent work has sought to amortize this cost through feed-forward prediction, 
demonstrating the inductive bias of 3D Gaussian representations can be combined with the generalization capacity of transformers to enable single-forward-pass reconstruction~\citep{charatan2024pixelsplat, chen2024mvsplat, zhang2024gs}.
Pose Splatter~\citep{goffinet2025pose} applies feed-forward 3DGS to multi-view animal behavior videos, using ``shape carving'' to initialize a voxel grid that is refined by a 3D U-Net. While Pose Splatter demonstrates accurate 3D reconstructions, it differs from \beastthree in key respects. First, Pose Splatter treats construction of a 3D latent representation as the end goal; \beastthree instead uses novel view synthesis as a pretext task to learn a general-purpose backbone whose features transfer directly to downstream tasks. Second, Pose Splatter requires foreground segmentation as a preprocessing step for shape carving, whereas \beastthree distills segmentation masks from a foundation model during training, so that inference requires only the raw multi-view images. 
Finally, \beastthree emphasizes multi-session training that generalizes better to new subjects.

\textbf{3D-aware pretraining from multi-view data.}
\label{sec:related:3d_pretrain}
A separate line of work uses 3D geometry not as an end in itself (like 3DGS) but as an inductive bias for learning transferable visual features.
RayZer~\citep{jiang2025rayzer} introduces ray-conditioned scene representations for novel view synthesis, using estimated Pl\"ucker ray coordinates~\citep{plucker1865xvii} to encode camera geometry alongside image features.
E-RayZer~\citep{zhao2025erayzer} extends this framework with an end-to-end architecture that jointly predicts camera poses and 3D Gaussians.
VGGT~\citep{wang2025vggt}
is a supervised model that predicts camera parameters, depth maps, and point clouds from a collection of images using alternating within- and across-view attention.
These methods demonstrate that explicitly reasoning about 3D geometry during pretraining produces features that outperform purely 2D pretraining on geometry-sensitive downstream tasks.
However, they are designed for general internet imagery where camera parameters are unknown and must be estimated jointly; models learn to do this by training on specialized datasets with dense camera coverage (often 10\texttt{+} views, e.g.,~\cite{zhou2018stereo,ling2024dl3dv}).
In laboratory settings, sparse scene coverage with 3-6 cameras is more common, and accurate camera calibration is readily available.
\beastthree exploits this by removing the camera parameter estimation branch entirely and instead conditions on ground-truth camera parameters.
This simplification reduces the number of trainable parameters, reduces the number of required views, and allows the model to focus capacity on learning appearance and geometry features.

\textbf{Self-supervised pretraining for behavior analysis.}
\label{sec:related:behavior_ssl}
Selfee~\citep{jia2022selfee} constructs composite frames from grayscale video sequences and applies standard contrastive learning techniques, demonstrating effectiveness on downstream tasks like action segmentation and anomaly detection.
Mueller et al.~\citep{mueller2025domain} and AnimalJEPA~\citep{zheng2024animal} adapt video-based joint-embedding predictive architectures to primate and mouse behavior classification, respectively.
These methods produce trajectory- or clip-level representations suited to action recognition but do not target multi-camera data.
Most closely related to our work, \beast~\citep{wang2026animal} combines masked autoencoding with temporal contrastive learning to pretrain a vision transformer on unlabeled videos from a single experimental setup, demonstrating improvements over general-purpose baselines on downstream neural encoding and pose estimation tasks.
However, \beast operates entirely in 2D image space and has no mechanism for reasoning about 3D geometry.
\beastthree extends this line of work by introducing an explicit 3D inductive bias: rather than reconstructing masked image patches, it reconstructs entire held-out viewpoints through differentiable 3D Gaussian splatting.
This formulation forces the representation to encode the full 3D structure of the subject---its shape, articulation, and appearance from arbitrary viewpoints---rather than merely interpolating 2D texture patterns.


%% file: contents/c3_methods.tex
\section{Method}
\label{sec:method}

\begin{figure*}[t!]
\begin{center}
  \includegraphics[width=\linewidth]{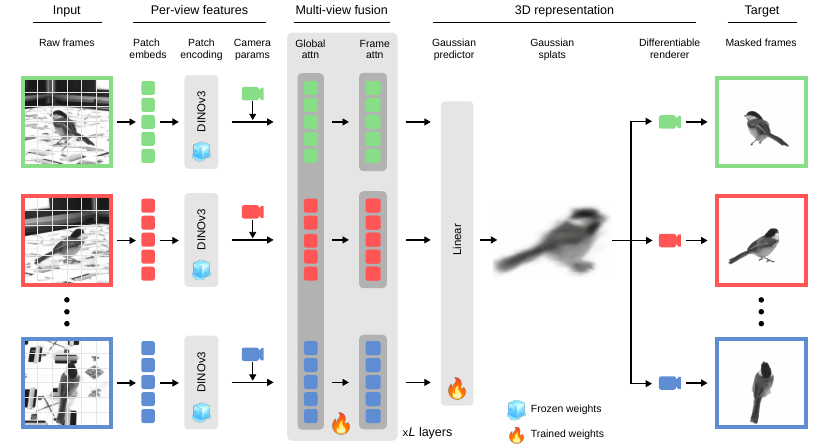}
\end{center} 
\caption{\textbf{\beastthree framework.} \beastthree is a masked autoencoder that uses 3D Gaussian splats as the intermediate representation. During training, one view is removed from the input and reconstructed through differentiable rendering of the 3D Gaussian splats inferred by the remaining views.
}
\label{fig:method}
\end{figure*}

We present \beastthree, a self-supervised framework that learns 3D visual representations from multi-view animal behavior videos.
Given a set of synchronized, calibrated camera views of a behaving animal, \beastthree is trained to reconstruct held-out viewpoints via differentiable 3D Gaussian splatting.
The resulting encoder captures rich geometric and appearance features that transfer effectively to downstream tasks.
Fig.~\ref{fig:method} provides an overview of our approach.

\subsection{Problem Setting}
\label{sec:method:setup}

$V$ synchronized cameras observe a subject from different viewpoints.
At each timestep we have a set of images $\{\mathbf{I}_v \in \mathbb{R}^{H \times W \times 3}\}_{v=1}^{V}$, the corresponding camera parameters, and foreground segmentation masks $\{\mathbf{M}_v \in \{0,1\}^{H \times W}\}_{v=1}^{V}$.
Our goal is to learn an encoder that maps images to a 3D scene representation, with no keypoint annotations.
At each training step, we randomly partition the $V$ views into a reference set $\mathcal{R}$ and a target set $\mathcal{T}$, with $|\mathcal{T}| \geq 1$.
The model observes only the reference views and camera parameters, predicts a 3D Gaussian representation of the scene, and renders images for all views.
The training signal comes from reconstruction errors from the held-out target views.

\subsection{Multi-view dataset construction}
\label{sec:method:dataset}


To construct training datasets we first uniformly sample one \textit{scene} (synchronized frames across cameras) per second from each recording. To isolate the animal from the background, \beastthree learns to predict per-pixel alpha values; by supervising this channel with ``ground truth'' segmentation masks during training, a separate segmentation model is not required at inference time. We generate these masks offline using SAM3~\citep{carion2025sam}: an initial bounding box per view is detected automatically from a simple text prompt (``mouse'', ``bird'', etc.) and propagated across subsequent frames. Running this process on every new recording at inference time would be impractical; SAM3 is computationally expensive, and mask propagation can fail on challenging frames (e.g., severe occlusions), requiring manual correction (see Fig.~\ref{fig:sam3_masks} for examples of SAM3 successes and failures). We therefore run it once carefully during dataset construction, using a lightweight GUI to correct propagation failures, and amortize this cost into \beastthree's trained weights. 


\subsection{Architecture}
\label{sec:method:arch}

\beastthree builds on the 3DGS framework of E-RayZer~\citep{zhao2025erayzer}, but replaces the learned image tokenizer and camera pose predictor with a frozen pretrained vision encoder (as in VGGT) and ground truth calibration data, respectively (Fig.~\ref{fig:method}).
These choices leverage the strong visual features of modern foundation models while exploiting the accurate camera geometry available in controlled lab settings.

\textbf{Image tokenization.}\label{sec:method:tokenizer}
We use a frozen DINOv3~\citep{simeoni2025dinov3} ViT-B/16 as the image encoder.
Each reference image $\mathbf{I}_v \in \mathbb{R}^{H \times W \times 3}$ is first normalized with ImageNet statistics and then passed through DINOv3 to obtain patch-level features $\mathbf{Z}_v \in \mathbb{R}^{N \times d}$, where $N$ is the number of spatial tokens and $d = 768$ is the feature dimension.
We retain the \texttt{[CLS]} and register tokens in the DINOv3 processing, allowing them to interact with the spatial tokens through self-attention; however, we discard the \texttt{[CLS]} and register tokens from the DINOv3 output, as these are not associated with any unique spatial patch, and would require a separate global-token pathway in the geometry transformer described below.
All DINOv3 parameters remain frozen throughout training.

\textbf{Camera tokenization.}
\label{sec:method:camera}
To provide \beastthree with geometric grounding, each pixel is associated with the ray that emanates from the camera center through that pixel's location in 3D space. We represent each such ray in Pl\"ucker coordinates~\citep{plucker1865xvii}, a compact 6D descriptor that jointly encodes the ray's direction in world space and its offset from the origin, and is invariant to position along the ray.
The resulting Pl\"ucker map $\boldsymbol{\Pi}_v \in \mathbb{R}^{H \times W \times 6}$ for view $v$ is tokenized into patches using a linear projection to produce camera tokens $\mathbf{P}_v \in \mathbb{R}^{N \times d}$.

\textbf{Token fusion.}
\label{sec:method:transformer}
For each reference view $v \in \mathcal{R}$, we augment both image and camera tokens by adding positional embeddings (fixed 2D sinusoidal positional embeddings passed through a separate two-layer MLP for each modality). We concatenate these augmented tokens $\mathbf{Z}_v^{'}$ and $\mathbf{P}_v^{'}$ along the feature dimension and fuse them with another simple two-layer MLP.

\textbf{Geometry transformer.} The fused tokens from all reference views are collected into a single sequence $\mathbf{F} = [\mathbf{F}_{v_1}; \ldots; \mathbf{F}_{v_{|\mathcal{R}|}}] \in \mathbb{R}^{|\mathcal{R}|N \times d}$ and processed by a pretrained VGGT geometry transformer~\citep{wang2025vggt}. The transformer consists of $L$ layers with self-attention alternating between two attention patterns: even layers apply frame attention, where each view attends only to its own tokens, while odd layers apply global attention across all views jointly.
This alternating strategy allows the model to interleave 2D appearance processing with 3D multi-view aggregation.

\textbf{3D Gaussian prediction.}
\label{sec:method:gaussian}
The output tokens from the geometry transformer are decoded into per-patch 3D Gaussian parameters via a linear head.
Each spatial token predicts one 3D Gaussian $\mathbf{g}_i$, yielding $|\mathcal{R}| \cdot N$ Gaussians in total. Each Gaussian is described by 27 parameters that define position, shape and view-dependent color information (Appendix~\ref{app:beast3d}).


\textbf{Differentiable rendering.}
\label{sec:method:render}
Given the predicted 3D Gaussians $\{\mathbf{g}_i\}_{i=1}^{|\mathcal{R}|N}$, we render images from both reference and target views using GSplat~\citep{ye2025gsplat}.
For each view, pixel colors are computed by accumulating Gaussian contributions along each viewing ray such that Gaussians closer to the camera contribute first, with each one partially occluding those behind it.
We apply an additional ``frustum constraint'' that culls Gaussians that fall outside the intersection of all views.

\subsection{Training}
\label{sec:method:training}

The training loss $\mathcal{L}$ is computed only on the held-out target views $\mathcal{T}$ and combines three terms: 
(1) a photometric loss $\mathcal{L}_{\ell_2}$ that penalizes pixel-level differences between the ground truth and rendered images; 
(2) a perceptual loss $\mathcal{L}_{\mathrm{perc}}$ that penalizes abstract feature-level differences between ground truth and rendered images; 
and (3) a mask loss $\mathcal{L}_{\mathrm{mask}}$ that penalizes differences between the rendered alpha channel and the SAM3 segmentation masks. The final loss is a weighted combination of each: 
\begin{equation}
    \mathcal{L} = \lambda_{\ell_2}\, \mathcal{L}_{\ell_2} + \lambda_{\mathrm{perc}}\, \mathcal{L}_{\mathrm{perc}} + \lambda_{\mathrm{mask}}\, \mathcal{L}_{\mathrm{mask}}.
    \label{eq:total_loss}
\end{equation}

In our experiments we set $\lambda_{\ell_2} = 1.0$, $\lambda_{\mathrm{perc}} = 0.2$, and $\lambda_{\mathrm{mask}} = 0.1$ across all datasets (Appendix~\ref{app:beast3d}).

We train \beastthree using the AdamW optimizer~\citep{loshchilov2017decoupled} with a base learning rate of $5 \times 10^{-5}$, weight decay of $0.05$, and a cosine learning rate schedule with 15\% warmup for 800 epochs.
Training uses bf16 mixed precision and is distributed across 8 GPUs. 
During each forward pass, we randomly sample $V_{\mathrm{ref}} = V - 1$ reference views and hold out $V_{\mathrm{tgt}} = 1$ target view.
The input image resolution is $256 \times 256$, producing $N = 256$ patch tokens per view.



%% file: contents/c4_experiments.tex

\section{Results}
\label{sec:results}

We first assess 3D reconstruction capabilities through a novel view synthesis task, which tests the models' ability to generate coherent 3D representations outside of the views it is prompted with.
We then demonstrate the versatility of \beastthree through two downstream neuro-behavioral tasks: pose estimation, a dense prediction task which assesses the model’s ability to extract 3D keypoints; and neural encoding, which challenges the model to extract features that predict patterns in neural activity.

\textbf{Datasets.} We evaluate \beastthree on four datasets spanning species, environments and camera configurations (Fig.~\ref{fig:fig1_baselines_compare}; Appendix~\ref{app:datasets}): (1) a head-fixed mouse recorded from six views (Cheese3D;~\cite{daruwalla2026cheese3d}); (2) a freely moving rat recorded from six views (Rat7M;~\cite{marshall2021continuous}); (3) a seed-caching chickadee recorded from six views~\citep{chettih2024barcoding}; (4) a human performing everyday tasks recorded from four views (Human3.6M~\cite{ionescu2013human3}).  Together, these datasets span both unconstrained naturalistic behavior and controlled lab settings.

\subsection{Novel view synthesis}
\label{sec:results:nvs}
Novel view synthesis (NVS) is the task of rendering a scene from a camera view that is withheld from the model at inference time (even though that same view may have been seen during training): for a given timestep, one or more views are held out and the model must reconstruct them from the remaining views alone.
This provides a direct window into the quality of the model's internal 3D representation: a model that generates geometrically consistent, photorealistic renderings of held-out views must have learned to infer the 3D structure of the scene rather than merely memorizing 2D appearance statistics.
We use NVS performance to benchmark \beastthree's 3D reconstruction capabilities against competing methods before turning to more specialized downstream tasks.

\begin{figure*}[t!]
\begin{center}
  \includegraphics[width=\linewidth]{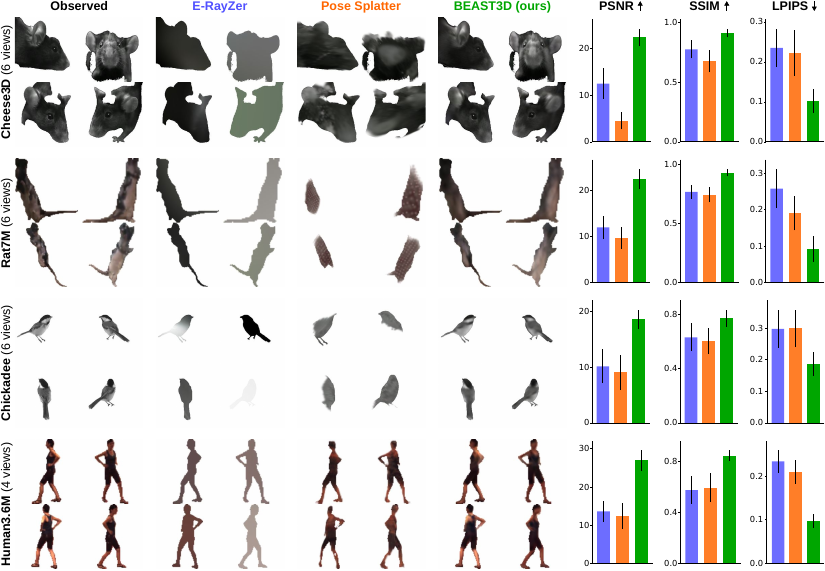}
\end{center} 
\caption{\textbf{\beastthree performs high-fidelity novel view synthesis.} 
\textit{Left}: example within-subject, held-out target views from each dataset and the corresponding reconstructions from E-RayZer, Pose Splatter, and \beastthree, each conditioned on the remaining views from the same timestep. Reconstructions are masked by the SAM3 outputs; within these masked regions, E-RayZer often produces empty renderings, indicating that its predicted Gaussians fail to populate the subject's location. 
\textit{Right}: per-dataset PSNR, SSIM, and LPIPS (described in text).
We assess significance with two-sided Wilcoxon signed-rank tests on paired image metrics, Holm-corrected across the nine contrasts within each dataset (three pairwise model contrasts times three metrics).
\beastthree significantly outperforms both baselines across all metrics and datasets at $p<10^{-10}$.}
\label{fig:novel_view_1}
\end{figure*}

\textbf{Baselines.}
We compare against three baselines: 
E-RayZer (zero-shot), the pretrained E-RayZer~\citep{zhao2025erayzer} model applied directly without any fine-tuning; 
E-RayZer (fine-tuned), the same model fine-tuned on each of our training datasets; 
and Pose Splatter~\citep{goffinet2025pose}, a feed-forward Gaussian splat model specifically designed for multi-view animal behavior recordings.
We note that Pose Splatter is extensively evaluated against per-scene optimization approaches---including 3DGS~\citep{kerbl20233d}, FSGS~\citep{zhu2024fsgs}, and GO~\citep{yang2024gaussianobject}---and was found to perform comparably or better overall; we therefore omit those baselines here. We also omit VGGT~\citep{wang2025vggt}, as it produces 3D point clouds but not the accompanying color, size, and opacity information for each point required to perform the NVS task.

\textbf{Evaluation.}
E-RayZer (fine-tuned), Pose Splatter, and \beastthree are each trained on data pooled across multiple individuals.
We evaluate in two complementary regimes: (1) within-subject, in which test frames are drawn from sessions included in training, but the specific test frames are held out; this matches the evaluation protocol of Pose Splatter. (2) Cross-subject, in which test frames are drawn from sessions and individuals not seen during training, assessing generalization to novel subjects.
In each setting we report three standard image quality metrics.
Peak signal-to-noise ratio (PSNR) quantifies pixel-level reconstruction performance
relative to the dynamic range of pixel intensities.
Structural similarity index measure (SSIM) captures perceptual similarity of two images by comparing luminance, contrast, and structure~\citep{wang2004image}.
Learned Perceptual Image Patch Similarity (LPIPS) measures perceptual dissimilarity between image patches using deep feature activations~\citep{zhang2018unreasonable}.
To enable fair comparisons across baselines, all metrics are computed over foreground pixels only, identified using SAM3 segmentation masks: E-RayZer reconstructs both foreground and background by default and must be masked accordingly, while Pose Splatter uses masks internally for shape carving.
\beastthree results evaluated without masks at inference time are reported in Appendix~\ref{app:beast3d}.

\textbf{Results.}
Qualitatively, \beastthree produces sharp reconstructions of held-out viewpoints that recover the subject's silhouette and fine appearance details across all four datasets (Fig.~\ref{fig:novel_view_1}; \hyperref[app:nvs]{Supplementary Videos}). With the ground truth masks applied, E-RayZer renderings clearly do not capture structure and appearance details of the subjects, while Pose Splatter outputs exhibit visible artifacts characteristic of shape-carving failures. 
Quantitatively, \beastthree improves metrics across all datasets in the within-subject evaluation setting (Fig.~\ref{fig:novel_view_1}; Tables~\ref{tab:nvs_ind_gt},~\ref{tab:nvs_ind_pred}). Across-subjects, \beastthree improves PSNR and LPIPS over the baselines on every dataset, and achieves the best SSIM on all but Chickadee, where E-RayZer is marginally better (Tables~\ref{tab:nvs_ood_gt},~\ref{tab:nvs_ood_pred}). E-RayZer fine-tuning yields no improvement, and occasionally degrades performance, relative to zero-shot (Appendix); we attribute this to its joint camera-pose estimation branch failing in the sparse view regime tested here (Fig.~\ref{fig:camera_pose}). 

\textbf{Ablations} (Tables~\ref{tab:nvs_ind_gt}--\ref{tab:nvs_ood_pred}). 
Following VGGT~\citep{wang2025vggt}, which employs a frozen DINOv2 encoder, we use DINOv3 as our per-view image encoder. Removing DINOv3 degrades NVS performance across all datasets except for Human3.6M, demonstrating the multi-view transformer is not by itself sufficient to learn high-quality geometric representations, and benefits from strong pretrained visual features.
Reinforcing this conclusion, the DINOv3 ablation \textit{substantially} degrades pose estimation performance (Fig.~\ref{fig:pose_ablation}), described in more detail in the next section.
We also ablate the frustum constraint, which culls Gaussians not contained within the intersection of all viewpoints. This constraint has minimal effect on most datasets but is critical for Chickadee, where the bird is cropped from large images using a per-frame bounding box.

\subsection{Pose Estimation}
Pose estimation underpins the quantitative study of behavior in neuroscience and ethology~\citep{datta2019computational, pereira2020quantifying}.
Although multi-view pose estimation is inherently 3D, most state-of-the-art pipelines process each view independently and triangulate 2D predictions into 3D as a separate post-processing step~\citep{nath2019using,karashchuk2021anipose}. 
In this work, we ask whether backbones that learn 3D structure during pretraining yield better pose estimators, particularly in the low-data regime typical of scientific applications.
We evaluate on each dataset introduced in the novel view synthesis task.

\begin{figure*}[t!]
\begin{center}
  \includegraphics[width=\linewidth]{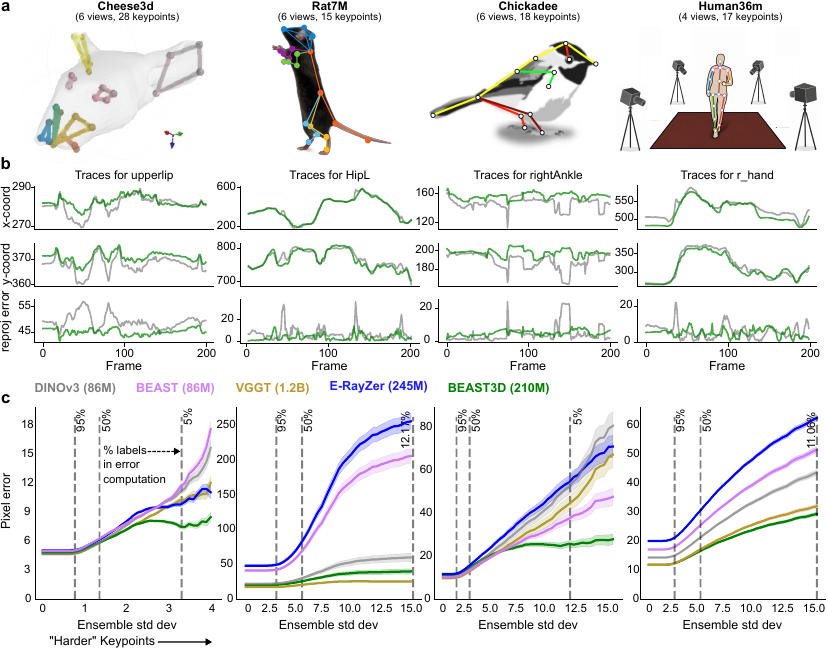}
\end{center} 
\caption{\textbf{\beastthree improves pose estimation.} 
\textbf{a:} Experimental setups and keypoint skeletons for all datasets. 
\textbf{b:} \textit{Top}: representative keypoint traces from a single view. \textit{Bottom}: corresponding 3D reprojection error for ViT-B DINOv3 (gray) and \beastthree (green). Because reprojection error leverages known camera geometry to measure agreement across views, it serves as a label-free proxy for prediction quality. Across all datasets, \beastthree traces are visibly smoother in time and incur substantially lower reprojection error.
\textbf{c:} Pixel error as a function of keypoint difficulty (lower is better), computed on frames from held-out test subjects~\citep{biderman2024lightning}.  
Dashed vertical lines indicate the fraction of data retained; error bands show s.e.m. across included keypoints. Results are shown for all datasets trained with subsets of 100 labeled instances.
}
\label{fig:pose}
\end{figure*}


\textbf{Models.} All pose estimators are trained with the Lightning Pose
package~\citep{biderman2024lightning,aharon2026lightning}. As a strong 2D
baseline we use a ViT-B/16 backbone pretrained with DINOv3~\citep{simeoni2025dinov3}; 
against this
baseline we evaluate four backbones:
\beast~\citep{wang2026animal}, a ViT-B/16 pretrained with masked autoencoding on single-view behavioral videos;
VGGT~\citep{wang2025vggt};
E-RayZer~\citep{zhao2025erayzer};
and \beastthree.

\textbf{Evaluation.} We adopt the limited-data evaluation protocol of the
original \beast paper~\citep{wang2026animal},
which is designed to reflect the annotation budgets of typical labs, where exhaustive labeling is impractical. For each backbone we train three models on random subsets of only 100 labeled instances and
evaluate on held-out test subjects.
Because pose estimation error varies substantially across keypoints---some are more inherently ambiguous or occluded---we report results using the difficulty-stratified error curves of ref.~\cite{biderman2024lightning}, which reveal how each method handles keypoints of varying difficulty rather than collapsing performance into a single number. 
``Difficulty'' is quantified by the standard deviation of predictions across models and seeds for an individual keypoint on an individual frame (ensemble standard deviation, e.s.d.; higher e.s.d. indicates higher disagreement among models). 
Each error curve plots mean pixel error (y-axis) against an e.s.d. threshold (x-axis), where each point includes only keypoints whose e.s.d. exceeds that threshold. The leftmost point thus summarizes all keypoints, while moving rightward progressively restricts to more ``difficult'' ones.

\textbf{Results.} We find that \beastthree achieves the best performance across nearly all datasets (Figs.~\ref{fig:pose},~\ref{fig:3d_pose}), with the exception of Rat7M where VGGT---a model nearly an order of magnitude larger---performs best.
Notably, \beastthree consistently outperforms \beast across all datasets, and across a range of label budgets (Fig.~\ref{fig:pose_frame_sweep}), demonstrating the benefit of multi-view pretraining. However, a striking finding is that multi-view models more broadly (VGGT and E-RayZer) do not consistently outperform single-view DINOv3 and \beast baselines, despite extensive fine-tuning and hyperparameter search (Appendix~\ref{app:pose}). This suggests that a multi-view architecture alone is insufficient to guarantee strong pose estimation performance; how a model is pretrained matters at least as much as its architectural capacity for 3D reasoning. \beastthree succeeds where other multi-view models fall short precisely because its novel view synthesis pretraining objective provides representations that transfer well to the pixel-level demands of pose estimation.


\subsection{Neural encoding}
A growing body of work probes the relationship between brain and behavior by predicting neural activity directly from behavior videos~\citep{musall2019single, stringer2019spontaneous, wang2026brain}. The standard approach relies on keypoints~\citep{syeda2024facemap, international2025reproducibility, international2025brain, daruwalla2026cheese3d}, potentially missing critical behavioral features that are not well captured by tracking a sparse set of points. \beast~\citep{wang2026brain} addressed this limitation by showing that learned 2D image representations outperform keypoints for neural encoding; however, these representations take the form of a single high-dimensional \texttt{CLS} token per frame, offering no spatial structure that would allow neuroscientists to interpret which aspects of the animal's appearance or posture drive neural activity. \beastthree's Gaussian splat representation is denser and more expressive than a sparse set of keypoints while remaining spatially grounded, making it more interpretable than opaque \texttt{CLS} tokens.

\begin{figure*}[t!]
\begin{center}
  \includegraphics[width=\linewidth]{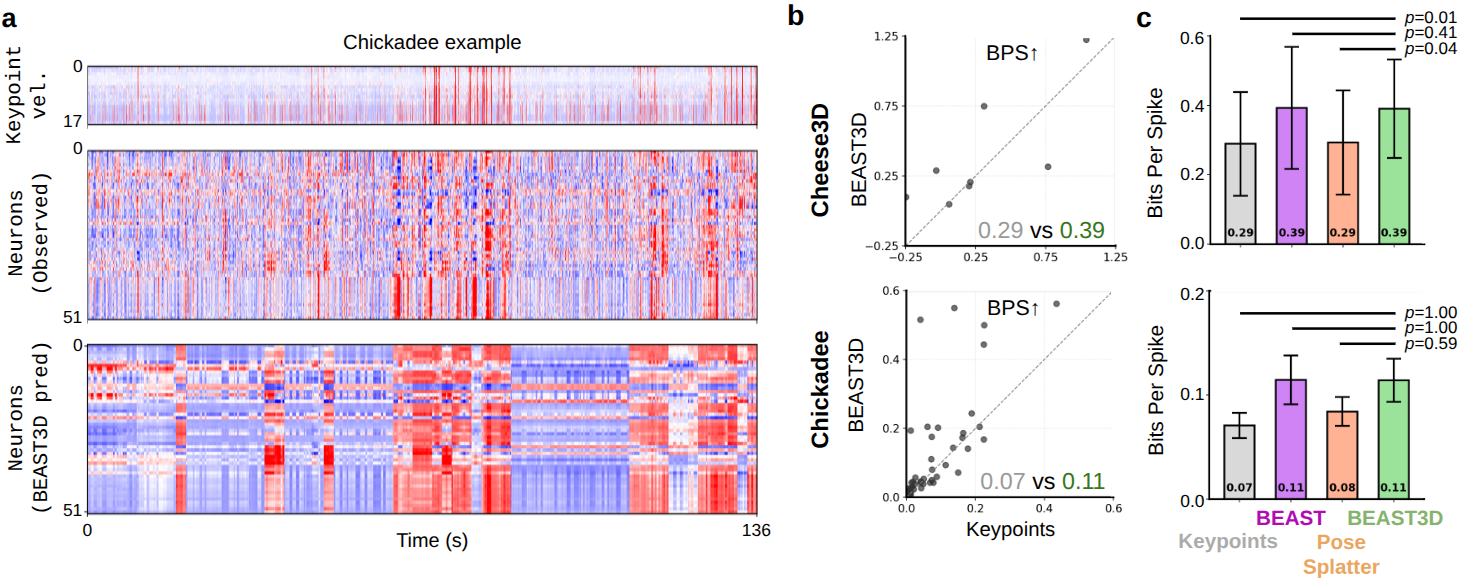}
\end{center}
\caption{\textbf{\beastthree~features improve neural encoding.} 
\textbf{a}:~Example session from Chickadee. \textit{Top}: z-scored 3D keypoint velocities. \textit{Middle}: observed neural activity. \textit{Bottom}: activity predicted from \beastthree~Gaussian splats on held-out timepoints. 
\textbf{b}: Per-neuron BPS for \beastthree~vs.\ keypoints; each dot is a neuron. 
Session-averaged BPS shown in bottom-right. 
\textbf{c}:~Average BPS across keypoints, \beast, Pose Splatter, and \beastthree, with error bars showing S.E.M.\ across neurons. 
We assess significance with two-sided Wilcoxon signed-rank tests on paired per-neuron $\Delta$BPS, Holm-corrected across the three contrasts within each dataset (\beastthree versus each of the baselines).
}
\label{fig:encoding}
\end{figure*}

\textbf{Datasets.} We evaluate neural encoding on two datasets: Cheese3D, with eight neurons recorded from the mouse facial motor nucleus during spontaneous behavior~\citep{daruwalla2026cheese3d}; 
and Chickadee, with 52 neurons recorded from hippocampus during seed-caching behavior~\citep{chettih2024barcoding}.
See Appendix~\ref{app:datasets} for details.

\textbf{Models.} We use a variety of feature representations for neural encoding: 
(1) sparse 3D keypoints from pose estimation; 
(2) \beast \texttt{CLS} tokens; 
(3) Pose Splatter Gaussian splats; 
and (4) \beastthree Gaussian splats. 
For the Gaussian splat models we keep the centroid of each Gaussian and discard the remaining parameters that are used for rendering (e.g., opacity, color, and scale), which we find do not contain meaningful information for this encoding task (Table~\ref{tab:ephys_appearance}).
For the keypoint- and splat-based representations we fit a PointNet transformer model~\citep{qi2017pointnet} applied per timepoint, followed by a spatiotemporal transformer that aggregates information over a fixed temporal window; for \beast we fit a temporal convolution network following~\citep{wang2026animal}.
See Appendix~\ref{app:encoding} for details.

\textbf{Evaluation.} We split neural activity from each session into non-overlapping 2 s chunks, and randomly assign each chunk to a train, val or test split (70\%/15\%/15\%). We tune model hyperparameters on the train/val splits, and report results on the test split using the Bits Per Spike (BPS) metric~\citep{pei2021neural}.

\textbf{Results.}
\beastthree~features outperform 3D keypoints for predicting neural activity in both datasets (Fig.~\ref{fig:encoding}), and per-neuron comparisons confirm that this gain is consistent across the population rather than driven by a small subset of well-predicted neurons. This demonstrates that dense 3D representations capture behavioral information that sparse keypoints discard. \beastthree also outperforms Pose Splatter, indicating that its splats capture more neurally-relevant behavioral structure. 
\beastthree~matches the performance of \beast~\texttt{CLS} tokens, recovering the predictive power of an opaque high-dimensional vector while preserving spatial structure. Because each Gaussian is anchored to a localized region of the subject, post-hoc analyses can ask which body parts or appearance details drive a given neuron's activity---a question that \texttt{CLS} tokens cannot answer.
Notably, these trends hold across both datasets despite substantial differences in species (mouse vs.\ bird) and brain region (facial motor nucleus vs.\ hippocampus).


%% file: contents/c5_conclusion.tex
\section{Conclusion}
\label{sec:conclusion}

We presented \beastthree, a self-supervised framework that learns dense 3D representations from multi-view animal behavior videos by reconstructing held-out viewpoints through differentiable Gaussian splatting. Across species and recording setups, we showed that \beastthree's representations support strong 3D awareness through novel view synthesis; improve standard pose estimation pipelines in the low-annotation regime; and provide a spatially-grounded, interpretable representation for predicting neural activity. In each setting, \beastthree outperforms general-purpose 3D computer vision models such as VGGT and E-RayZer, which are designed for the dense-view, uncalibrated regime and struggle in the sparse-view laboratory setting. \beastthree also generalizes more reliably across subjects than Pose Splatter, a model tailored to multi-view animal recordings.

A current limitation of \beastthree is its compute requirements: pretraining on a single dataset takes roughly 32 hours on 8 A100 GPUs, placing it out of reach for many labs (see Appendix~\ref{app:compute} for details on inference compute). Identifying architectural choices that can reduce parameter count while preserving representation quality is an important avenue for future work. The framework naturally accommodates joint training across multiple datasets, which could yield a single backbone that labs use out-of-the-box without dataset-specific pretraining. In a first experiment, a single model trained jointly on three datasets matched the per-dataset models on held-out frames of training subjects but generalized worse to new subjects (Appendix~\ref{app:nvs:joint}; Tables~\ref{tab:nvs_ind_gt} and~\ref{tab:nvs_ood_gt}), so more careful training protocols will be needed to realize this direction. 

Several extensions of \beastthree suggest themselves as natural next steps. On the representation side, \beastthree does not currently produce a canonical shape space: each timestep's Gaussian splats are predicted independently, with no explicit correspondence across frames or individuals. Designing a mechanism for a canonical body representation that is shared across frames---much as in the mesh-based literature~\citep{kanazawa2018learning,zuffi20173d,wu2023magicpony,li2024learning,zhong2026moremouse}---while remaining species-agnostic and template-free, is a promising direction for future work. A related but more constrained extension would fit an articulated skeleton or mesh model to \beastthree's dense per-frame observations, though this would require species-specific body templates~\citep{zuffi2019three,bohnslav2023armo,klibaite2025mapping}.
On the application side, \beastthree's fixed-coordinate geometry is naturally suited to egocentric alignment and depth-sensitive posture analysis, which raises the possibility of using \beastthree embeddings for automated behavior discovery~\citep{wiltschko2015mapping,hsu2021b,luxem2022identifying,weinreb2024keypoint,goffinet2025pose}; our datasets lack ground-truth ethograms, however, so we present this as a motivated direction for future evaluation. A further extension is to multi-animal social settings, where prior work has found 3D representations to be particularly informative for analyzing interactions~\citep{rajasegaran2023benefits,klibaite2025mapping,martini2026bigmaq}.
Together, these directions promise to lower the barrier to adoption and bring self-supervised 3D representation learning into broader use across the behavioral neuroscience community.

%% file: contents/c6_acks.tex
\textbf{Acknowledgments and Disclosure of Funding}

\small{
We thank Qitao Zhao for discussions about the E-RayZer model, and Jack Goffinet for discussions about the Pose Splatter model. 
This work was supported by the following: 
Gatsby Charitable Foundation GAT3708, 
NIH 1R50NS145433,
NIH U19NS123716, 
NSF 1707398, 
the National Science Foundation and DoD OUSD (R\&E) under Cooperative Agreement PHY-2229929 (The NSF AI Institute for Artificial and Natural Intelligence), 
Simons Foundation, 
Wellcome Trust 216324, 
and funds from Zuckerman Institute Team Science.
The authors declare no competing interests.
}

%% file: appendix/main.tex
\begin{center}
    {\LARGE \textbf{Supplementary Material}}\\[0.5em]
    \rule{0.55\linewidth}{0.4pt}\\[0.7em]
    {\large \beastthree: Animal behavioral analysis and neural encoding \\[0.15em]
    from multi-view video via Gaussian splatting}
\end{center}
\vspace{1.5em}


\tableofcontents


\vspace{10pt}
\section{Datasets}
\label{app:datasets}

For all datasets we define a set of ``In-Distribution'' (InD) sessions used for training and a separate set of ``Out-of-Distribution'' (OOD) sessions comprising \textit{new subjects} for evaluation. 
Table~\ref{tab:dataset_stats} documents the number of sessions, subjects and \beastthree training/evaluation frames for each dataset; Appendix~\ref{app:dataset_construction} details how these frames are selected.
For the pose estimation task, training frames are drawn exclusively from InD sessions and test frames exclusively from OOD sessions. 

\subsection{Cheese3D} 

\paragraph{Behavior data.} A head-fixed mouse behaves spontaneously, captured by six cameras at 100 Hz~\citep{daruwalla2026cheese3d}. Frame sizes are 640 $\times$ 512 pixels.

For the pose estimation task, we did not have the required labeled dataset where each keypoint is labeled across all views for a given instance in time. We instead constructed a pseudo-labeled dataset from an initial set of 665 instances where each keypoint is labeled in a subset of views depending on anatomical visibility. Using this data, we trained an ensemble of three single-view transformer pose estimation models \citep{aharon2026lightning} using a DINOv2-pretrained ViT-B backbone (Fig.~\ref{fig:pose_svt}; step 2). 

\textit{Ensemble inference.} We ran inference with 
the ensemble
across a set of full-length videos. For each time point, view and keypoint, we computed the median across the ensemble as our 2D prediction.
Because training labels covered only a subset of views per keypoint, predictions in unlabeled views were systematically unreliable and received low likelihoods.

\textit{Frame filtering and triangulation.} We applied a confidence filter requiring that at least two views report a median likelihood exceeding 0.6 for a given keypoint to contribute to triangulation. Keypoints for which no view ever exceeded this threshold within a given recording session---arising from anatomical occlusion or a complete absence of training annotations across all views---were excluded from the per-frame acceptance criterion and appear as \texttt{NaN} in the final pseudo-labeled output. A frame was retained if all eligible keypoints were confidently predicted in at least two views simultaneously. For retained frames, we performed multi-view triangulation using camera parameters to recover 3D keypoint locations. These 3D estimates were then reprojected into all six views, producing geometrically consistent 2D pseudo-labels even in cameras that lacked direct confident predictions.

\textit{Frame selection.} To obtain a compact yet pose-diverse pseudo-labeled set, we applied $k$-means clustering in the space of flattened 3D keypoints for each session. 
The frame closest to each cluster centroid was selected as the representative.
We targeted 50--55 representative frames per session.

\textit{Reprojection error quality filtering.} As a final quality control step, we computed a per-frame reprojection error as the mean Euclidean distance between the original ensemble median 2D predictions and the triangulation-derived reprojected labels, averaged over all confident keypoints and views. We discarded the worst 25\% of frames by this metric.
Pseudo-labels are provided for all six views per retained frame, with \texttt{NaN} entries where triangulation was not possible. The final pose estimation training/test sets consist of 450/150 instances, respectively. 

\paragraph{Neural data.} A 32-channel single-shank silicon probe was inserted into the facial motor nucleus. The probe was coated with lipophilic dyes to reveal the probe track post hoc. Recordings began at least 15 min after probe insertion to ensure recording stability. Voltage signals were acquired at 30 kHz. After the recording, single electrical pulses were delivered to all sites on the probe to induce facial movements to verify probe placement location. For this specific recording session, ear, whisker pad, nose and mouth movements were observed following electric stimulation.

Spikes were sorted into 8 well-isolated single units and binned at the downsampled video frame rate (50~Hz). The full $\sim$20~min session (61{,}697 frames) was segmented into non-overlapping 2.0~s windows of 100 frames each, yielding 616 windows. Windows were shuffled with a fixed seed and partitioned into 70\% / 15\% / 15\% train / val / test splits (431 / 92 / 93 windows). All 8 units were retained without firing-rate thresholding. 

\subsection{Rat7M} 
A single freely moving rat behaves spontaneously in a circular arena, captured by six cameras at 120 Hz~\citep{marshall2021continuous}. Frame sizes are 1328 $\times$ 1048 pixels.

Rat7M is a marker-based motion capture dataset with a large number of ground truth pose labels spanning multiple subjects. We retain 15 of the original keypoints in each view, omitting those tracking the head stage of the rat.
Rather than retain the 30,000+ labeled instances in the original dataset, we chose to curate a smaller subset of instances matched in size with our other datasets. For each session we first filter out any instances with missing data. We then filter out potentially problematic points using skeleton distances (\texttt{ElbowR}-\texttt{ElbowL} distance in $[40, 60]$; \texttt{KneeR}-\texttt{ShinR} and \texttt{KneeL}-\texttt{ShinL} distances in $[10, 1000]$).
We then run $k$-means clustering on the remaining 3D poses (using 100 clusters per session) and select one example per cluster. Finally, we performed manual inspection of the resulting labels and excluded any instances where ground truth keypoints were incorrect due to camera syncing issues. The pose estimation training/test sets consist of 455/177 instances, respectively.

The dataset is available at \url{https://doi.org/10.6084/m9.figshare.c.5295370} under the CC-BY 4.0 license.

\subsection{Chickadee} 

\paragraph{Behavior data.}
Freely moving chickadees engage in seed caching behavior in a large arena, captured by six cameras at 60 Hz~\citep{chettih2024barcoding}. Frame sizes vary by view but are approximately 3000 $\times$ 1500 pixels. Given the small size of the bird relative to the arena, we produced a set of cropped videos for model training. Using previously collected pose estimation labels, we trained a detector network on full resolution frames downsampled to 256 $\times$ 256 pixels to localize the bird within each frame. We computed a bounding box from the pose estimation output, and use these cropped videos for training all downstream models.

For the pose estimation task, we created a cropped dataset using the ground truth labels to define a bounding box around the bird, and reshaped the cropped frames to 320 $\times$ 320 pixels. Eighteen keypoints on the chickadee's body are labeled in each view. The training/test sets consist of 433/143 instances, respectively.

\paragraph{Neural data.} Large-scale silicon-probe recordings of one chickadee yielded 132 spike-sorted units across a multi-hour free-behavior session~\citep{chettih2024barcoding}. Spikes were binned at the video frame rate (60~Hz). To focus the encoding analysis on a behaviorally rich and neurally active window, we performed a sliding-window search over the full recording for the contiguous 15-minute interval that maximized the number of units firing at $\geq$1~Hz; this interval (54{,}000 frames at 60~Hz, in which 53 of 132 units exceed 1~Hz mean firing rate) is used for all subsequent encoding analyses. After cutting the videos and spike trains to this range, we segmented the spike trains into non-overlapping 2.0~s windows of 120 frames each (450 windows total), shuffled with a fixed seed, and split 70\% / 15\% / 15\% into train / val / test (315 / 67 / 68 windows). For neural encoding, we additionally drop units whose mean spike count per training window is below 2 (i.e.\ $<$1~Hz), retaining 52 units; this filter is computed on training windows only and applied to all three splits to avoid data leakage. 

\subsection{Human3.6M}
Human subjects perform a range of everyday activities captured by four synchronized cameras at 50~Hz~\citep{ionescu2011latent, ionescu2013human3}. Frame sizes are 1000 $\times$ 1002 pixels.

Human3.6M is a marker-based motion capture dataset with ground truth pose labels spanning multiple subjects and 15 activity types (actions 2--16). We retain 17 keypoints in each view. 
Rather than retain the 3.6M labeled instances in the original dataset, similar to Rat7M we chose to curate a smaller subset of instances to benchmark the pose estimation models. 
Only the first subaction of each action is used; subaction variants and action~13 of subject~9 are excluded due to corrupted 3D coordinates.
For each subject-action pair, we first filter out any instances with missing 3D 
keypoint data. We then run $k$-means clustering on the remaining 3D poses using 15 clusters and select the single instance closest to each cluster center, yielding up to 15 representative frames per session. 
We additionally require that selected frame indices are separated by at least 3 frames. 
The pose estimation training/test sets consist of 1125/425 instances, respectively.

The dataset is available at \url{http://vision.imar.ro/human3.6m} under the Human3.6M license.

\begin{table}[t]
\centering
\small
\caption{Dataset statistics for In-Distribution (InD) and Out-of-Distribution (OOD) splits.}
\label{tab:dataset_stats}
\resizebox{\textwidth}{!}{
\begin{tabular}{lcccccccc}
\toprule
& \multicolumn{4}{c}{In-Distribution} & \multicolumn{3}{c}{Out-of-Distribution} \\
\cmidrule(lr){2-5} \cmidrule(lr){6-8}
Dataset & Subjects & Sessions & Train frames & Test frames & Subjects & Sessions & Test frames \\
\midrule
Cheese3D~\citep{daruwalla2026cheese3d}    & $6$ & $11$ & $44370$ & $8382$ & $2$ & $4$ & $13890$ \\
Rat7M~\citep{marshall2021continuous}       & $3$ & $5$ & $53970$ & $7650$ & $2$ & $2$ & $21588$ \\
Chickadee~\citep{chettih2024barcoding}   & $6$ & $12$ & $64728$ & $7650$ & $2$ & $4$ & $21558$ \\
Human3.6M~\citep{ionescu2013human3}   & $5$ & $75$ & $78596$ & $15472$ & $2$ & $30$ & $27660$ \\
\bottomrule
\end{tabular}
}
\end{table}

\section{Dataset construction}
\label{app:dataset_construction}

All datasets are processed by a single, configurable pipeline that turns raw multi-view recordings into the calibrated \textit{(frame, mask, camera)} tuples consumed by \beastthree. The pipeline runs five stages in order: \emph{cut}, \emph{downsample}, \emph{segment}, \emph{assemble} and \emph{resize}.
Each stage writes its output to disk and the next stage picks it up automatically, so individual steps can be re-run without redoing earlier work. The same pipeline is invoked three times per dataset, with three different configurations, to produce the pretraining set, the in-distribution (InD) test set, and the out-of-distribution (OOD) test set.

\subsection{Processing pipeline}
\paragraph{Cut.} Many of the source recordings are very long (tens of minutes to several hours) and captured at 50--120~Hz, which is far denser than what is needed for self-supervised pretraining of a static-frame model. The optional \emph{cut} step trims each video to a user-specified inclusive frame range using \texttt{ffmpeg}'s frame-accurate \texttt{trim} filter, so that downstream steps see only the portion of the recording that contains the behavior of interest. Whenever bounding-box CSVs accompany the videos, they are filtered and re-indexed to the same range so that bbox-frame alignment is preserved. 

\paragraph{Downsample.} The \emph{downsample} step then resamples each video to a target frame rate via \texttt{ffmpeg -vf fps=...} and re-encodes with libx264 (CRF~18, audio dropped). Both \textit{cut} and \textit{downsample} steps process video files in parallel through a process pool to amortize cluster I/O.

\paragraph{Segment.} For each downsampled video we run SAM3~\citep{carion2025sam} in tracking mode with a dataset-specific text prompt (e.g.\ \texttt{mouse} for Cheese3D, \texttt{bird} for Chickadee) to obtain a per-frame binary foreground mask. 
When SAM3's text grounding fails for a particular video---usually because the subject is small, occluded in the first frame, or visually atypical---the failure is recorded in \texttt{failed\_videos.json}. We developed a small Gradio app to manually click two corners of a bounding box on the first frame of each failed video; a follow-up \emph{retry} stage then re-runs SAM3 on those videos with the labeled boxes supplied, which bypasses text grounding and resolves all remaining failures.

\begin{figure}[t]
\centering
\includegraphics[width=\linewidth]{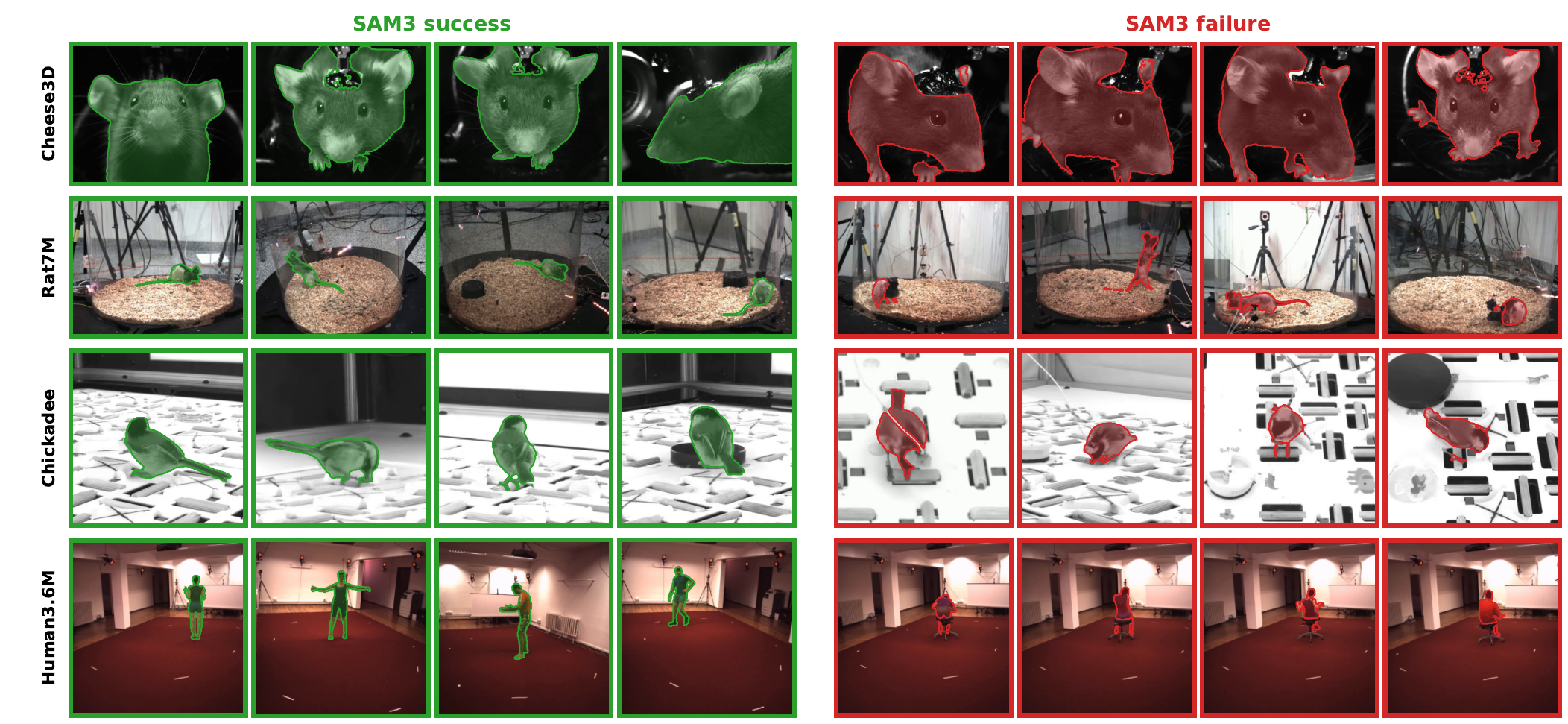}
\caption{\textbf{SAM3 mask quality.} Representative SAM3 foreground masks (colored overlay and contour) for each dataset: successes (\textit{green}, left) and failures (\textit{red}, right). Failures are mostly truncated, fragmented, or leaking masks caused by occlusion or contact with scene objects (feeders, cables, a chair).}
\label{fig:sam3_masks}
\end{figure}

Figure~\ref{fig:sam3_masks} shows representative SAM3 masks for all four datasets. In the large majority of frames SAM3 produces a single connected mask that tightly follows the subject's silhouette. The typical failure mode involves occlusion: the mask is truncated or split into fragments where the subject touches or is partially hidden by scene objects (feeders, perches, cables, a chair), or it leaks onto the occluding object itself. We also observe that masks are qualitatively more reliable on Human3.6M than on the animal datasets (mouse, rat, and bird), likely because human imagery is far better represented than laboratory animal footage in the training data of the segmentation foundation model.

\paragraph{Assemble.} The \emph{assemble} stage groups videos by session and camera, 
loads the corresponding intrinsics and extrinsics from the calibration file (Anipose \texttt{toml} or DeepLabCut \texttt{pickle}, depending on the dataset), and writes a single self-contained directory per session of the form \texttt{<session>/<cam>/img\{idx\}.png} together with a corresponding \texttt{img\{idx\}.npy} that holds the per-frame camera parameters (and bounding box, when available). The matching SAM3 mask is copied alongside as \texttt{mask\{idx\}.png}. 


\paragraph{Resize.} A final \emph{resize} stage rescales the assembled images and masks so that the shorter side is 256~pixels, matching the input resolution expected by the DINOv3 ViT-B/16 backbone used in \beastthree (Appendix~\ref{app:beast3d}). Calibration intrinsics are scaled by the same factor.

\subsection{Dataset splits}
\paragraph{Pretraining set.} The pretraining configuration uses the InD sessions of each dataset and aggressively downsamples each video to a low effective frame rate---1~Hz for Cheese3D and Chickadee, and a comparable rate for Rat7M and Human3.6M---under the assumption that consecutive seconds are visually near-redundant for the purpose of single-frame multi-view geometry learning. After the cut, downsample, segment, assemble, and resize stages, every retained frame becomes a training example with up to six paired views, foreground masks, and calibrated cameras. This procedure produces the training-frame counts reported in Table~\ref{tab:dataset_stats}.

\paragraph{OOD test set.} The OOD test set is constructed by re-running the exact same pipeline on a held-out set of sessions from \emph{new subjects} that never appear during pretraining. Apart from pointing at a different input directory, the configuration is identical to the pretraining one---same target frame rate, same SAM3 prompt, same calibration loader, same resize---so the resulting frames are statistically comparable to the training frames in resolution, masking quality, and temporal density. This isolates the effect of subject identity on novel view synthesis and pose estimation performance.

\paragraph{InD test set.} The InD test set is more delicate: it must come from the same sessions as pretraining, yet contain frames that are guaranteed to be unseen during training. We achieve this by re-running the downsample stage on the same InD videos with a \emph{phase-shifted} frame selector. Concretely, if the source video is captured at $f_{\text{src}}$~Hz and the pretraining target is $f_{\text{tgt}}$~Hz, the pretraining pipeline keeps source frames at indices $\{0, S, 2S, \ldots\}$ with stride $S = \mathrm{round}(f_{\text{src}} / f_{\text{tgt}})$. The InD test pipeline shifts this lattice by an offset $K$ with $1 \le K < S$, keeping source frames $\{K, K+S, K+2S, \ldots\}$. We use $K = S/2$ (e.g.\ $K=50$ for Cheese3D's $f_{\text{src}}=100$~Hz, $f_{\text{tgt}}=1$~Hz pretraining setup), which places each test frame roughly half a sampling interval away from the nearest training frame. Because $1 \le K < S$, the training and InD test frame sets are guaranteed to be disjoint at the source-frame level, while still being drawn from the same underlying distribution of behavior, lighting, and subjects. 

\section{\beastthree}
\label{app:beast3d}

For each timestep, for a dataset with $V$ views, we have access to a set of images $\{\mathbf{I}_v \in \mathbb{R}^{H \times W \times 3}\}_{v=1}^{V}$, the corresponding camera-to-world transformations $\{\mathbf{T}_v \in \mathrm{SE}(3)\}_{v=1}^{V}$ and intrinsic parameters $\{f_x^v, f_y^v, c_x^v, c_y^v\}_{v=1}^{V}$, and foreground segmentation masks $\{\mathbf{M}_v \in \{0,1\}^{H \times W}\}_{v=1}^{V}$ via SAM3.

At each training step, we randomly partition the $V$ views into a reference set $\mathcal{R}$ and a target set $\mathcal{T}$, with $|\mathcal{T}| \geq 1$. Our goal is to learn an encoder that maps reference views and their camera parameters to a 3D scene representation which is capable of reconstructing the target views.

\subsection{Architecture details}

\paragraph{Image tokenization.}
We use a frozen DINOv3~\citep{simeoni2025dinov3} ViT-B/16 as the image encoder.
Each reference image $\mathbf{I}_v \in \mathbb{R}^{H \times W \times 3}$ is first normalized with ImageNet statistics and then passed through DINOv3 to obtain patch-level features:
\begin{equation}
    \mathbf{Z}_v = \mathrm{DINOv3}(\mathbf{I}_v) \in \mathbb{R}^{N \times d}, \quad N = \left(\frac{H}{p}\right)^2,
\end{equation}
where $p = 16$ is the patch size, $N$ the number of spatial tokens, and $d = 768$ the feature dimension.
We discard the \texttt{[CLS]} token and any register tokens, retaining only the spatial patch tokens.
All DINOv3 parameters remain frozen throughout training.

\paragraph{Camera tokenization.}
For a pixel at location $(u, w)$ in view $v$ with intrinsics $(f_x, f_y, c_x, c_y)$ and camera-to-world matrix $\mathbf{T}_v = [\mathbf{R}_v \mid \mathbf{t}_v]$, we first compute the ray direction in world coordinates:
\begin{equation}
    \mathbf{d}_{u,w} = \mathbf{R}_v \cdot \begin{pmatrix} (u + 0.5 - c_x)/f_x \\ (w + 0.5 - c_y)/f_y \\ 1 \end{pmatrix}, \quad \hat{\mathbf{d}}_{u,w} = \frac{\mathbf{d}_{u,w}}{\|\mathbf{d}_{u,w}\|},
\end{equation}
and the ray origin $\mathbf{o}_v = \mathbf{t}_v$.
The 6D Pl\"ucker representation is then:
\begin{equation}
    \boldsymbol{\pi}_{u,w} = [\mathbf{o}_v \times \hat{\mathbf{d}}_{u,w};\; \hat{\mathbf{d}}_{u,w}] \in \mathbb{R}^6,
    \label{eq:plucker}
\end{equation}
where the first three components are the moment vector and the last three are the direction.

The Pl\"ucker coordinate map $\boldsymbol{\Pi}_v \in \mathbb{R}^{H \times W \times 6}$ for view $v$ is tokenized into patches using a linear projection (analogous to the image tokenizer in DINOv3) to produce camera tokens $\mathbf{P}_v \in \mathbb{R}^{N \times d}$.

\paragraph{Token fusion.}
For each reference view $v \in \mathcal{R}$, we augment both image and camera tokens with positional embeddings, which are fixed 2D sinusoidal positional embeddings passed through a separate two-layer MLP for each modality. Each MLP consists of a linear layer, SiLU activation, and another linear layer.
We concatenate the augmented tokens $\mathbf{Z}_v^{'}$ and $\mathbf{P}_v^{'}$ along the feature dimension and fuse them with a two-layer MLP:
\begin{equation}
    \mathbf{F}_v = \mathrm{MLP}_{\mathrm{fuse}}\big(\, [\mathbf{Z}_v^{'} \,;\, \mathbf{P}_v^{'}] \,\big) \in \mathbb{R}^{N \times d},
    \label{eq:fusion}
\end{equation}
where $\mathrm{MLP}_{\mathrm{fuse}}: \mathbb{R}^{2d} \to \mathbb{R}^{d}$ consists of layer normalization, a linear layer, SiLU activation, and another linear layer.

\paragraph{Geometry transformer.} The fused tokens from all reference views are collected into a single sequence $\mathbf{F} = [\mathbf{F}_{v_1}; \ldots; \mathbf{F}_{v_{|\mathcal{R}|}}] \in \mathbb{R}^{|\mathcal{R}|N \times d}$ and processed by a geometry transformer.
Following VGGT~\citep{wang2025vggt}, the transformer consists of $L$ layers with QK-normalized multi-head self-attention~\citep{henry2020query} that alternate between two attention patterns:
\begin{itemize}
    \item \textbf{Frame attention} (even layers): attention is computed independently within each view, allowing the model to reason about local 2D structure.
    Tokens are reshaped to $(|\mathcal{R}| \cdot B) \times N \times d$, where $B$ is the batch size, so that each view attends only to its own tokens.
    \item \textbf{Global attention} (odd layers): attention spans all views jointly, enabling cross-view reasoning about 3D geometry.
    Tokens are reshaped to $B \times (|\mathcal{R}| \cdot N) \times d$ so that each token attends across all reference views.
\end{itemize}
This alternating strategy allows the model to interleave 2D appearance processing with 3D multi-view aggregation, balancing computational efficiency with geometric expressiveness.
Layer weights are initialized with a depth-dependent standard deviation $\sigma_\ell = 0.02 / \sqrt{2(\ell + 1)}$ to stabilize training of deep transformers~\cite{wang2025vggt}.

\paragraph{3D Gaussian prediction.}
The output tokens from the geometry transformer are decoded into per-patch 3D Gaussian parameters via a linear head.
Each spatial token predicts one 3D Gaussian, yielding $|\mathcal{R}| \cdot N$ Gaussians in total.
Specifically, the decoder predicts the following attributes per Gaussian:
\begin{equation}
    \mathbf{g}_i = (\Delta\mathbf{x}_i,\; \mathbf{c}_i,\; \mathbf{s}_i,\; \mathbf{q}_i,\; \alpha_i),
\end{equation}
where $\Delta\mathbf{x}_i \in \mathbb{R}^3$ is a position offset, $\mathbf{c}_i \in \mathbb{R}^{n_{\mathrm{SH}} \times 3}$ are spherical harmonic (SH) coefficients with $n_{\mathrm{SH}} = (\ell_{\max}+1)^2$ and $\ell_{\max} = 3$ for view-dependent color, $\mathbf{s}_i \in \mathbb{R}^3$ is the log-scale, $\mathbf{q}_i \in \mathbb{R}^4$ is the rotation quaternion, and $\alpha_i \in \mathbb{R}$ is the pre-sigmoid opacity.
The log-scale is clamped to $[s_{\min}, s_{\max}]$ for numerical stability.

\paragraph{Hard pixel alignment.}
To anchor the predicted Gaussians in 3D space, we apply a hard pixel-alignment strategy~\citep{jiang2025rayzer}.
The position offset $\Delta\mathbf{x}_i$ is first mapped to a depth value $\delta_i \in [\delta_{\mathrm{near}}, \delta_{\mathrm{far}}]$ via a linear range function.
The final 3D position is obtained by marching along the corresponding camera ray:
\begin{equation}
    \mathbf{x}_i = \mathbf{o}_v + \delta_i \cdot \hat{\mathbf{d}}_i,
    \label{eq:pixelalign}
\end{equation}
where $\mathbf{o}_v$ is the ray origin and $\hat{\mathbf{d}}_i$ is the normalized ray direction at the spatial location of token $i$.
This constrains each Gaussian to lie on the ray cast from its corresponding pixel, which provides a strong geometric prior and improves convergence.

\subsection{Loss details}

The \beastthree loss is the weighted sum of three losses: photometric, perceptual, and mask losses.

\paragraph{Photometric loss.}
We use a foreground-weighted mean squared error between the rendered image $\hat{\mathbf{I}}_v$ and an augmented ground-truth target image $\mathbf{I}_v^*$ for each target view $v \in \mathcal{T}$:
\begin{equation}
    \mathcal{L}_{\ell_2} = \frac{1}{|\mathcal{T}|}\sum_{v \in \mathcal{T}} \left[\, w_{\mathrm{fg}} \cdot \frac{\sum (\hat{\mathbf{I}}_v - \mathbf{I}_v^*)^2 \odot \mathbf{M}_v}{\sum \mathbf{M}_v} + \frac{\sum (\hat{\mathbf{I}}_v - \mathbf{I}_v^*)^2 \odot (1 - \mathbf{M}_v)}{\sum (1 - \mathbf{M}_v)} \,\right],
\end{equation}
where $w_{\mathrm{fg}} = 5$ up-weights the foreground region to focus learning on the subject rather than the (augmented) background.
During training, a random background color $\mathbf{b} \sim \mathrm{Uniform}([0,1]^3)$ is composited into the initial ground truth image $\mathbf{I}_v$ via $\mathbf{I}_v^* \leftarrow \mathbf{I}_v \odot \mathbf{M}_v + \mathbf{b} \odot (1 - \mathbf{M}_v)$ to prevent the model from memorizing a fixed background.

\paragraph{Perceptual loss.}
Following E-RayZer~\citep{zhao2025erayzer}, we add a VGG-based perceptual loss~\citep{johnson2016perceptual} that extracts multi-scale features from a pretrained VGG-19 network:
\begin{equation}
    \mathcal{L}_{\mathrm{perc}} = \frac{1}{|\mathcal{T}|}\sum_{v \in \mathcal{T}} \sum_{l=0}^{5} w_l \cdot \|\phi_l(\hat{\mathbf{I}}_v) - \phi_l(\mathbf{I}_v^*)\|_1,
\end{equation}
where $\{\phi_l\}$ are the feature maps at selected VGG layers and $\{w_l\}$ are normalization weights.
This loss encourages perceptually plausible reconstructions and reduces blurriness.

\paragraph{Mask loss.}
We supervise the rendered alpha channel $\hat{\mathbf{A}}_v$ (accumulated opacity from Gaussian splatting) against the ground-truth foreground mask:
\begin{equation}
    \mathcal{L}_{\mathrm{mask}} = \frac{1}{|\mathcal{T}|}\sum_{v \in \mathcal{T}} \|\hat{\mathbf{A}}_v - \mathbf{M}_v\|_2^2.
\end{equation}
We use MSE rather than binary cross-entropy to avoid excessively large gradients in background regions, which typically occupy the majority of the image.

\section{Baselines} \label{app:baselines}

\subsection{VGGT}
VGGT~\citep{wang2025vggt} is a feed-forward 3D foundation model that, given an unordered set of multi-view images, jointly predicts per-pixel world-space points (with a per-point confidence), per-view camera extrinsics, and per-view intrinsics in a single forward pass. We use VGGT strictly as an off-the-shelf reference: we do \emph{not} fine-tune it on any of our animal datasets and instead use the public pretrained checkpoint released by the authors\footnote{\url{https://github.com/facebookresearch/vggt}}. At inference we resize each input view to $224\times224$ pixels (VGGT's training resolution), run a single forward pass over all available views with bf16 autocast, and recover (i)~per-view camera extrinsics and intrinsics, and (ii)~a single shared point cloud by concatenating the per-pixel world points across views. To suppress low-confidence background floaters we remove points whose confidence falls in the bottom 50\% of all predicted points; the remaining points retain their per-pixel RGB colors and form the VGGT point cloud used in Fig.~\ref{fig:fig1_baselines_compare}.
No alignment to ground-truth cameras is performed for the point cloud itself; renders shown in Fig.~\ref{fig:fig1_baselines_compare} are rasterized either through the VGGT-predicted cameras directly (Cheese3D, Rat7M, Chickadee) or through the ground-truth cameras (Human3.6M) after a per-frame Procrustes alignment of the ground truth camera pose into VGGT's predicted world frame (the same alignment used for the NVS evaluation in Appendix~\ref{app:nvs}).

\subsection{E-RayZer}
E-RayZer~\citep{zhao2025erayzer} is a 3D-aware transformer that consumes multi-view images and predicts a set of per-pixel 3D Gaussians, which are then rendered into target views via Gaussian splatting\footnote{\url{https://github.com/QitaoZhao/E-RayZer}}. Unlike VGGT, we treat E-RayZer as a fine-tunable baseline: we initialize from the public pretrained checkpoint and continue self-supervised pretraining on each of our datasets independently. Concretely, the model first tokenizes each input image with a ViT-style image tokenizer, runs a VGGT-style alternating frame/global attention encoder, and decodes per-view camera extrinsics and intrinsics from a learned camera token. These \emph{predicted} per-view cameras are converted into Pl\"ucker ray embeddings, fused with the image tokens via an MLP, processed by a geometry transformer, and decoded into per-pixel Gaussian parameters with hard pixel alignment along the corresponding camera ray; the same predicted cameras are then used as the rendering cameras for Gaussian splatting, both during pretraining and for novel-view synthesis at inference.

The four main differences from \beastthree are: (i)~E-RayZer predicts its own cameras rather than consuming the calibrated ground truth cameras as input, and renders through those predicted cameras, whereas \beastthree consumes and renders through ground truth cameras; (ii)~E-RayZer uses a simple patch tokenizer before sending patch embeddings into the geometry transformer, whereas \beastthree uses a frozen DINOv3 backbone as a much more powerful patch tokenizer, similar to VGGT; (iii)~E-RayZer does not enforce the frustum constraint that \beastthree uses to anchor Gaussians within the visible volume of the reference cameras; and (iv)~E-RayZer is trained to reconstruct its inputs with no foreground supervision, whereas \beastthree adds a foreground-weighted reconstruction term and a mask MSE term that supervise the rendered alpha against the SAM3 segmentation mask (Appendix~\ref{app:beast3d}). The first three differences shift what \emph{geometry} the model can recover, while the loss difference shifts what part of the image the model is encouraged to fit.

\begin{figure}[h]
\centering
\includegraphics[width=\linewidth]{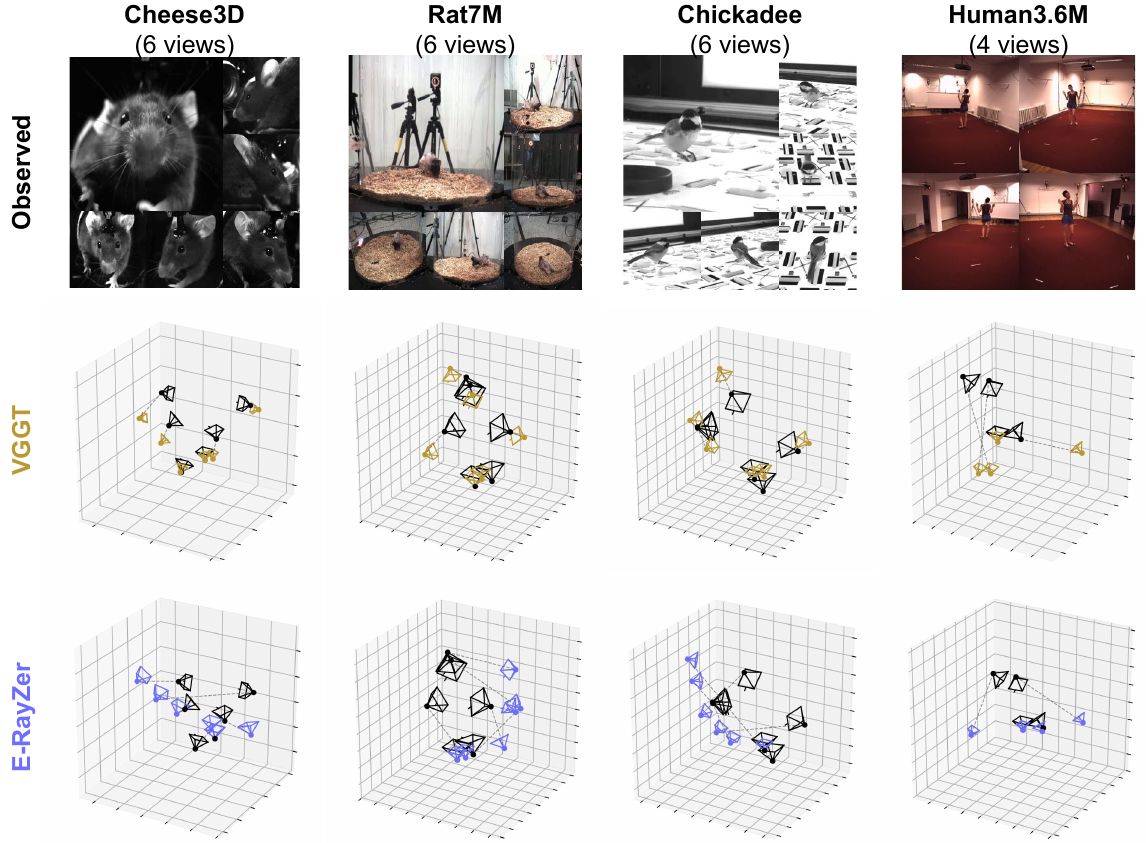}
\caption{\textbf{Camera pose prediction collapses in the sparse-view regime.}
\textit{Top:} representative input views.
\textit{Middle:} VGGT-predicted cameras (\textit{colored}) paired with ground truth cameras (\textit{black}) via dashed lines.
\textit{Bottom:} E-RayZer-predicted cameras, also paired with ground truth via dashed lines.
VGGT's predictions stay close to the ground truth, but deviate more strongly on the Human3.6M dataset which only has four views. E-RayZer---which learns camera poses fully unsupervised---produces poses that are nearly uncorrelated with the true cameras under our 4--6 widely separated views.}
\label{fig:camera_pose}
\end{figure}

A practical caveat is that E-RayZer was originally pretrained with 10 views per scene split into 5 reference + 5 target, a regime that assumes dense, highly overlapping coverage of the same object (as in scene-scale datasets such as DL3DV~\citep{ling2024dl3dv}). Animal behavior recordings in our lab datasets only provide 4--6 widely separated cameras whose fields of view share a small common volume, and in this sparse-camera regime E-RayZer's pose prediction collapses --- the predicted cameras are essentially uncorrelated with the true poses (Fig.~\ref{fig:camera_pose}), which propagates to the Pl\"ucker ray embeddings and prevents the geometry transformer from converging.

\paragraph{Fine-tuning.} E-RayZer was fine-tuned for 200 epochs with AdamW (lr $5\times10^{-5}$, weight decay $0.05$) at a global batch size of 256, retaining E-RayZer's original loss weights.

\subsection{Pose Splatter}

Pose Splatter (PS)~\citep{goffinet2025pose} is a feed-forward multi-view 3D Gaussian splatting model designed for multi-view animal datasets. Given calibrated cameras and per-view foreground masks, PS constructs a coarse voxel occupancy volume via multi-view silhouette intersection (``shape carving''), refines this representation with a 3D U-Net, and decodes occupied voxels into Gaussian parameters for differentiable rendering. Unlike generalizable reconstruction models, PS is trained per dataset.

PS differs from \beastthree along several design dimensions that affect how 3D structure is inferred. 
(i) PS uses foreground masks as direct inputs to shape carving at inference time, whereas \beastthree uses foreground masks only as supervision during training and predicts geometry from unmasked RGB images at inference time. 
(ii) PS initializes geometry via multi-view silhouette intersection before network refinement, introducing an explicit geometric constraint induced by mask consistency and camera alignment. \beastthree instead predicts Gaussians directly from learned multi-view image features conditioned on camera rays. 
(iii) PS is trained to optimize photometric reconstruction of rendered views, while \beastthree uses held-out-view reconstruction with an additional perceptual loss to combat vision transformer patch artifacts. 
These differences reflect distinct inductive biases in how geometry and appearance are represented and optimized.

\paragraph{Training.} We use the public implementation\footnote{\url{https://github.com/jackgoffinet/pose-splatter}} and train one model from scratch per dataset, following the provided preprocessing pipeline to produce geometry-normalized inputs with the same SAM3-generated foreground masks as those used in \beastthree training. Models are trained for 50 epochs using Adam with a learning rate of $1\times10^{-4}$ and batch size 1. The optimizer is used without weight decay, following the official implementation. We otherwise follow the official dataset-specific training configurations, including a voxel grid resolution of $64^3$ for Cheese3D, Rat7M, and Human3.6M, and $112^3$ for Chickadee, with dataset-specific cropping volumes aligned to the canonical object frame. 

\paragraph{Evaluation.} For evaluation, we compare two shape carving protocols. The \emph{original} protocol follows the PS evaluation setup, where all available views contribute to shape carving. The \emph{leave-one-out} (LOO) protocol matches our held-out target-view setting: for target view $v$, the shape carver uses only the remaining $V\!-\!1$ views at evaluation time, and the model is scored by rendering view $v$. Table~\ref{tab:ps_protocol} reports this comparison on the InD split with ground truth mask scoring. The original protocol improves PSNR on all datasets. We use the LOO protocol in Tables~\ref{tab:nvs_ind_gt}--\ref{tab:nvs_ood_pred}.

\begin{table}[h]
\centering
\small
\caption{\textbf{Effect of Pose Splatter carving protocol on InD GT-mask NVS.}
Original uses all available views for shape carving; LOO excludes the rendered target view from the carver at evaluation time.}
\label{tab:ps_protocol}
\resizebox{\textwidth}{!}{%
\begin{tabular}{lccc ccc ccc ccc}
\toprule
& \multicolumn{3}{c}{Cheese3D}
& \multicolumn{3}{c}{Rat7M}
& \multicolumn{3}{c}{Chickadee}
& \multicolumn{3}{c}{Human3.6M} \\
\cmidrule(lr){2-4} \cmidrule(lr){5-7} \cmidrule(lr){8-10} \cmidrule(lr){11-13}
Protocol
& PSNR$\uparrow$ & SSIM$\uparrow$ & LPIPS$\downarrow$
& PSNR$\uparrow$ & SSIM$\uparrow$ & LPIPS$\downarrow$
& PSNR$\uparrow$ & SSIM$\uparrow$ & LPIPS$\downarrow$
& PSNR$\uparrow$ & SSIM$\uparrow$ & LPIPS$\downarrow$ \\
\midrule
Original 
& 13.703 & 0.600 & 0.420
& 8.360 & 0.699 & 0.194
& 12.366 & 0.630 & 0.309
& 10.020 & 0.742 & 0.187 \\
LOO
& 12.443 & 0.594 & 0.419
& 4.519 & 0.673 & 0.223
& 9.064 & 0.603 & 0.299
& 9.648 & 0.743 & 0.191 \\
\bottomrule
\end{tabular}
}
\end{table}

\subsection{\beast}
\beast~\citep{wang2026animal} is a single-view, self-supervised pretraining procedure for animal behavior video. It pretrains a ViT-B/16 backbone with masked autoencoding (MAE)~\citep{he2022masked} on individual frames, treating each camera view as an independent image with no cross-view fusion or camera input. We pretrain one backbone per dataset using the public implementation\footnote{\url{https://github.com/paninski-lab/beast}}, with the default recipe (mask ratio 0.75, AdamW optimizer, cosine learning rate schedule, 800 epochs); the optional InfoNCE contrastive term is disabled so that pretraining reduces to pure single-view MAE. Because \beast does not see 3D structure or cameras, it serves both as the natural 2D baseline against \beastthree and as a fine-tunable backbone for the single-view pose estimation models in Appendix~\ref{app:pose}.

\section{Novel view synthesis}
\label{app:nvs}

We report full novel view synthesis (NVS) results across all four datasets
(Cheese3D, Rat7M, Chickadee, Human3.6M) and six methods
(E-RayZer~\citep{zhao2025erayzer} zero-shot (ZS), E-RayZer fine-tuned (FT)
per dataset, Pose Splatter~\citep{goffinet2025pose}, \beastthree without
the frustum constraint, \beastthree without the DINOv3 backbone, and the
full \beastthree). For each
method--dataset pair we evaluate two regimes that match
Section~\ref{sec:results:nvs}: the In-Distribution Test (InD Test) split,
in which held-out frames come from training subjects, and the
Out-of-Distribution Test (OOD Test) split, in which test frames come from
unseen subjects. We also evaluate each setting under two masking
conventions: foreground masked with the ``ground truth'' SAM3 segmentation
mask (Tables~\ref{tab:nvs_ind_gt} and \ref{tab:nvs_ood_gt}), and
foreground masked with the rendered alpha channel of each method
(Tables~\ref{tab:nvs_ind_pred} and \ref{tab:nvs_ood_pred}). The latter
additionally lets us report foreground-segmentation intersection over
union (IoU) against the ``ground truth'' mask. Numbers are mean $\pm$
standard deviation across all evaluation frames; the best result per
dataset and metric is highlighted in bold (ties at the displayed precision
are bolded jointly). E-RayZer~(ZS) and E-RayZer~(FT) do not produce a
confident foreground alpha and are therefore omitted from the
predicted-mask tables. Tables~\ref{tab:nvs_ind_gt} and~\ref{tab:nvs_ood_gt}
additionally include \beastthree~(joint), a single model trained on
Cheese3D, Rat7M and Human3.6M and evaluated only on those datasets
(Appendix~\ref{app:nvs:joint}).

We provide example InD Test videos for each dataset, which show the synchronized camera views (\textit{top left}), distilled foreground segmentation mask (\textit{top right}), predicted 3D Gaussian splats for each view (\textit{bottom left}), and a free-viewpoint orbit of the reconstructed subject demonstrating novel view synthesis capabilities (\textit{bottom right}):
\href{https://drive.google.com/file/d/1HkEutekV7lNBg4TSji6c-5L8jB47Niq0/view?usp=sharing}{Cheese3D},
\href{https://drive.google.com/file/d/1kMEZCXY-EvxcI55DC_ouTTn0NSjGQ_uw/view?usp=sharing}{Rat7M},
\href{https://drive.google.com/file/d/1h8YuLAj_uvpmKmaB8JER4ir52-xWiVlY/view?usp=sharing}{Chickadee},
\href{https://drive.google.com/file/d/1iZwJyGmwFgArqI4ECfBoMJaxMa13fBUM/view?usp=sharing}{Human3.6M}.

\subsection{Joint training across datasets}
\label{app:nvs:joint}

The other \beastthree models in this paper are trained separately on each dataset, whereas \beastthree~(joint) is a single model trained on the union of the Cheese3D, Rat7M and Human3.6M InD training sets (Table~\ref{tab:dataset_stats}), which together contain 14 subjects from three species. We leave Chickadee out of this initial experiment to keep the setup simple, since its frames are cropped from a large arena with a per-frame bounding box (Appendix~\ref{app:datasets}). The camera rigs differ (six cameras for Cheese3D and Rat7M, four for Human3.6M), so each mini-batch is drawn from a single dataset and uses all of that dataset's views. Batches from the three datasets are shuffled together, so each dataset is sampled in proportion to its size. All other settings follow the per-dataset recipe in Section~\ref{sec:method:training}, except that the first 261 of the 800 epochs ran on 4 GPUs (global batch size 128) instead of 8 (global batch size 256) because of compute availability. We evaluate the last saved checkpoint, at epoch 781.

Tables~\ref{tab:nvs_ind_gt} and~\ref{tab:nvs_ood_gt} report \beastthree~(joint) next to the per-dataset \beastthree. We compare the two with two-sided Wilcoxon signed-rank tests on paired image metrics, Holm-corrected across the six contrasts within each dataset (two splits times three metrics). All six contrasts are significant at $p<10^{-20}$ on every dataset. Since thousands of paired frames make even small differences significant, we also check whether each difference is consistent across recording sessions. On the InD Test split, the joint model improves SSIM and LPIPS on all three datasets and in nearly every session. Its PSNR is on par with the per-dataset models, marginally higher on average but not consistently across sessions.

On the OOD Test split, where all subjects are new, the joint model has lower PSNR than the per-dataset models on all three datasets and in every session. On Cheese3D it still improves SSIM and LPIPS in all four OOD sessions, while on Rat7M and Human3.6M it is worse on all three metrics. Rat7M has only two OOD sessions, and most of its gap comes from one of them.

The InD Test results show that one set of weights can reconstruct head-fixed mice, freely moving rats and humans recorded with different camera rigs as well as the per-dataset models do, with consistent gains in SSIM and LPIPS. We view this as encouraging for a cross-subject, cross-species 3D foundation model for animal behavior. Pooling does not, however, improve generalization to new individuals, plausibly because it adds species and camera rigs but not more individuals per species (three to six per dataset). We expect better cross-subject reconstruction to require multi-view training data with many more subjects per species.

\clearpage
\newpage

\input{appendix/nvs_tables}

\clearpage
\newpage

\section{Inference compute cost}
\label{app:compute}

\begin{figure}[h]
\centering
\includegraphics[width=\linewidth]{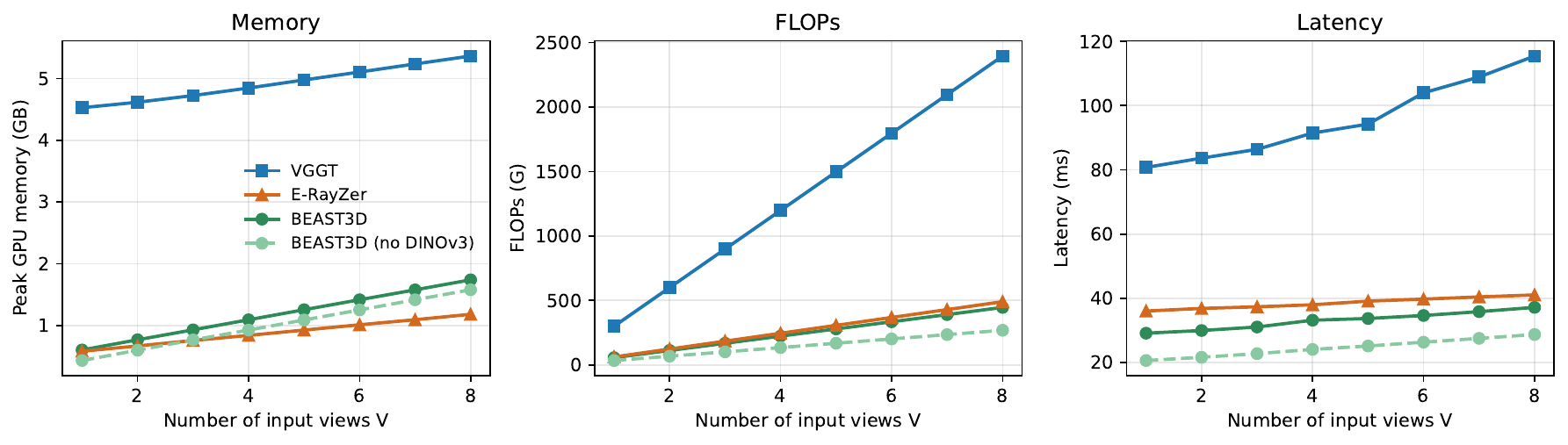}
\caption{\textbf{Inference compute cost vs.~number of input views.} Peak GPU memory (left), FLOPs (middle), and median latency over 20 timed iterations after 5 warmup iterations (right) for VGGT, E-RayZer, and \beastthree (with and without DINOv3), swept over $V\in\{1,\ldots,8\}$ at batch size 1, $256\times256$ input. Default deployment config, single GPU, bfloat16 autocast.}
\label{fig:compute_benchmark}
\end{figure}

We benchmark inference cost of \beastthree, VGGT, and E-RayZer as the number of input views grows (Fig.~\ref{fig:compute_benchmark}). All models run in their default deployment configuration on a single GPU with synthetic $256\times256$ inputs in bfloat16. VGGT is the most expensive along every axis (roughly $5\times$ the FLOPs and latency of the lightweight models). \beastthree and E-RayZer have comparable compute and memory profiles, with the DINOv3 frontend in \beastthree adding a roughly constant per-view memory and latency overhead.

\section{Pose estimation} \label{app:pose}
\subsection{Model training}
\label{sec:model_training}
We evaluate five backbone architectures spanning both single-view
and 3D-aware multi-view models. All models share a common training framework built on Lightning Pose~\citep{ biderman2024lightning,aharon2026lightning}, with architecture-specific modifications described below. For all models, backbone features are transformed into per-view keypoint heatmaps using a lightweight head consisting of transposed convolutions, following ViTPose~\citep{xu2022vitpose}, which demonstrated that ViT backbones are sufficiently expressive that simple linear/deconvolution heads are sufficient for strong pose estimation performance. Standard DLC-style augmentations~\citep{mathis2018deeplabcut}---including rotation, scaling, and cropping---are applied during training. Images are resized to $256 \times 256$ pixels and heatmaps are downsampled by a factor of 4, yielding $64 \times 64$ target heatmaps per keypoint per view. All models are trained for 300 epochs with the Adam optimizer and a multi-step learning rate schedule that decays the rate by a factor of $0.5$ at epochs 150, 200, and 250. We employ a two-stage training strategy: backbone weights are frozen for the first 20 epochs to allow the randomly initialized prediction head to adapt to the backbone's feature space, after which the entire network is fine-tuned end-to-end. We use 95\% of labeled frames for training and 5\% for validation, and sample 100 labeled frames across all views per dataset for all experiments. We monitor the validation loss every 5 epochs and retain the checkpoint with the lowest validation loss as the final model.

\begin{figure*}[t!]
\begin{center}
  \includegraphics[width=\linewidth]{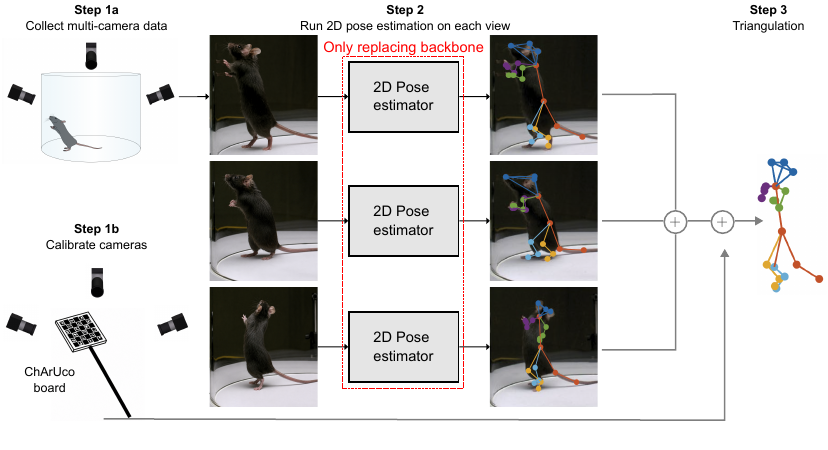}
\end{center} 
\caption{\textbf{Pose estimation pipeline for single-view heatmap models.} 
Step 1: Collect synchronized multi-camera data and calibrate cameras using a ChArUco board. Step 2: Run 2D pose estimation independently on each view, sweeping the backbone across single-view heatmap models for comparison. Step 3: Triangulate the per-view 2D predictions into 3D keypoints using the calibrated camera parameters. Some images modified from~\cite{dunn2021geometric}.
}
\label{fig:pose_svt}
\end{figure*}

\paragraph{Single-view heatmap models.}
The first group of methods processes each camera view independently in the same way (Fig.~\ref{fig:pose_svt}) through a shared view-agnostic backbone followed by the heatmap head. We consider two backbones in this category:
\begin{itemize}
    \item \textbf{ViT-B DINOv3}: A ViT-B pretrained with DINOv3 self-supervised learning~\citep{simeoni2025dinov3}, which learns semantically rich patch-level representations without explicit supervision.
    \item \textbf{\beast}: A ViT-B pretrained using the \beast self-supervised pretraining procedure on the target dataset~\citep{wang2026animal}, then fine-tuned for pose estimation. This two-stage approach provides dataset-specific feature initialization, and was previously shown to outperform convolution-based architectures (i.e.\ ResNet-50) as well as other transformer-based architectures (e.g., DINOv1 and DINOv2).
\end{itemize}

In both single-view models, per-view heatmaps are predicted independently and the final 2D keypoint locations are obtained via soft-argmax on each heatmap. These models use a training batch size of 8.
\paragraph{Multi-view 3D-aware models.}
The second group of methods explicitly leverages multi-view geometry by fusing information across camera views within the model architecture. We consider three architectures:
\begin{itemize}
    \item \textbf{VGGT}: A ViT-L backbone pretrained with DINOv2, paired with an alternating attention aggregator that performs iterative frame-level and global-level cross-view attention to fuse multi-view features. VGGT is the largest model in our comparison.
    \item \textbf{E-RayZer}: A 3D-aware transformer that uses a ViT-B backbone with VGGT-style cross-view attention layers to aggregate information across views. The backbone is initialized from dataset-specific E-RayZer pretrained weights.
    \item \textbf{\beastthree}: Extends the \beast pretraining paradigm to the 3D setting, using a ViT-B backbone with cross-view transformer layers. The backbone is initialized from dataset-specific \beastthree pretrained weights.
\end{itemize}
E-RayZer and \beastthree use a training batch size of 8 and require camera calibration parameters as additional input. For VGGT, we reduce the training batch size to 4 and employ gradient accumulation over 2 steps (effective batch size of 8) to accommodate the model's memory requirements on a single GPU.

\subsection{Hyperparameter selection}
\label{sec:hyperparameter_selection}
For each (backbone, dataset) pair, we perform an independent learning rate
sweep over five candidates spanning two orders of magnitude:
$\{1\mathrm{e}{-5},\, 5\mathrm{e}{-5},\, 1\mathrm{e}{-4},\, 5\mathrm{e}{-4},\, 1\mathrm{e}{-3}\}$.
We train three models per learning rate with different random seeds
controlling both the data split and heatmap head weight initialization, record the minimum validation heatmap MSE achieved during training for each seed, and average across seeds. The learning rate minimizing this averaged metric is used for all subsequent training and evaluation. Selecting per pair ensures that no method is penalized by a shared hyperparameter choice.

\subsection{Additional results}
\begin{figure*}[h!]
\begin{center}
  \includegraphics[width=\linewidth]{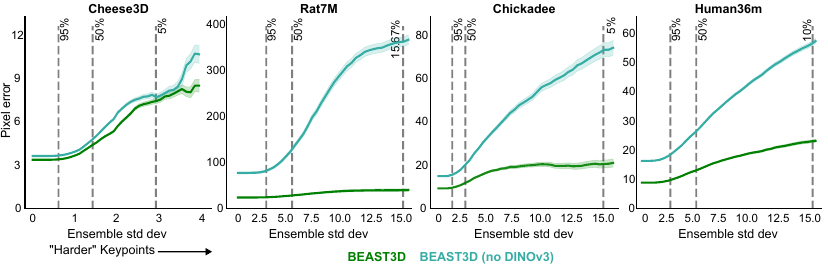}
\end{center} 
\caption{\textbf{Pose estimation results with DINOv3 ablation.} Figure conventions as in Fig.~\ref{fig:pose}.
}
\label{fig:pose_ablation}
\end{figure*}

\textbf{Ablations.} We report pose estimation results using the \beastthree ablation without the per-view DINOv3 encoder, which produces substantially worse results (Fig.~\ref{fig:pose_ablation}). 

\begin{figure*}[h!]
\begin{center}
  \includegraphics[width=\linewidth]{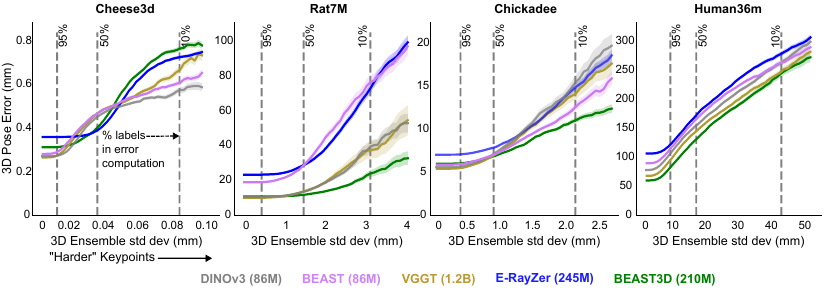}
\end{center} 
\caption{\textbf{3D pose estimation results.} Figure conventions as in Fig.~\ref{fig:pose}.
}
\label{fig:3d_pose}
\end{figure*}

\textbf{3D error.} The pose estimation results in Fig.~\ref{fig:pose} report 2D pixel error, aggregated across all views; we additionally report 3D error in physical space measured in millimeters (Fig.~\ref{fig:3d_pose}).

\begin{figure*}[h!]
\begin{center}
  \includegraphics[width=\linewidth]{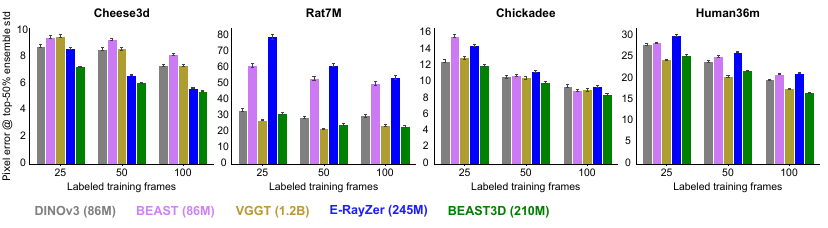}
\end{center} 
\caption{\textbf{Few-shot pose estimation across labeling budgets.} Pixel error at the 50\% ensemble standard deviation threshold (lower is better) for backbones fine-tuned on 25, 50, or 100 labeled training frames, evaluated on held-out OOD test subjects. Bars show mean$\pm$s.e.m. across three seeds (distinct train/val splits). Backbone parameter counts are shown in the legend.}
\label{fig:pose_frame_sweep}
\end{figure*}

\textbf{Few-shot pose estimation.} All previous pose estimation results fine-tune backbones on 100 labeled frames (Figs.~\ref{fig:pose},~\ref{fig:pose_ablation},~\ref{fig:3d_pose}). To further test the few-shot abilities of the different backbones, we also fine-tuned on 25 and 50 labeled frames, again using 3 seeds each (different train/val splits per seed), with the 25-frame set nested within the 50-frame set and the 50-frame set nested within the 100-frame set for a given seed. We repeated the error-versus-uncertainty quantification on the held-out test sets for each label budget (Fig.~\ref{fig:pose_frame_sweep}), and find that \beastthree maintains its general superiority across all budgets.

\clearpage
\newpage
\section{Neural encoding}
\label{app:encoding}

We assess whether the 3D representations recovered by \beastthree carry information about simultaneously recorded neural activity by training a simple encoding model that maps a window of 3D structure to a window of per-neuron firing rates. The same model architecture and training recipe are used for both Cheese3D and Chickadee datasets; only the inputs differ. Table~\ref{tab:ephys_stats} summarizes the two electrophysiology datasets used in this section.

\begin{table}[h]
\centering
\small
\caption{Summary of the two electrophysiology datasets used for neural encoding. ``Spike windows'' counts the non-overlapping 2.0~s segments within the video used as encoding trials.}
\label{tab:ephys_stats}
\begin{tabular}{lll}
\toprule
& \textbf{Cheese3D} & \textbf{Chickadee} \\
\midrule
Subject / area & mouse facial motor nucleus~\citep{daruwalla2026cheese3d} & chickadee hippocampus~\citep{chettih2024barcoding} \\
Views & 6 & 6 \\
Video FPS & 50~Hz (downsampled) & 60~Hz (native) \\
Video frames & 61{,}697 ($\approx$20.6~min) & 54{,}000 (15~min) \\
Spike binning & 50~Hz (matched to video) & 60~Hz (matched to video) \\
Window length & 2.0~s = 100 frames & 2.0~s = 120 frames \\
Train / val / test windows & 431 / 92 / 93 (total 616) & 315 / 67 / 68 (total 450) \\
3D keypoints per frame & 28 & 18 \\
Point-cloud points per frame & 8{,}192 & 1{,}024 \\
Sorted units (raw) & 8 & 132 \\
Filter policy & none & 1~Hz \\
Units used for encoding & 8 & 52 \\
\bottomrule
\end{tabular}
\end{table}

\paragraph{Per-frame 3D point clouds.}
For each session, we run the pretrained \beastthree model in a held-out per-frame inference pass. For every video frame and every reference view, the geometry transformer predicts one pixel-aligned 3D Gaussian per spatial token (Eq.~\ref{eq:pixelalign}); we only keep the centroid of each Gaussian and discard the rotation, opacity, scale, and spherical-harmonic parameters that are only used for rendering. We retain only Gaussians whose predicted opacity exceeds $0.05$ \emph{and} whose corresponding pixel falls inside the SAM3 foreground mask, which removes background floaters and emphasizes the subject. For Cheese3D we additionally clip points to a fixed axis-aligned bounding box around the head and keep at most the top 8{,}192 points per frame ranked by opacity; for Chickadee no bounding box is applied and the full foreground point set is kept. The resulting per-frame point clouds, together with their integer frame indices, are concatenated across the session and serve as the visual input to the encoding model.

\paragraph{Encoding model.}
The encoding model is intentionally lightweight and input-agnostic so that it does not itself learn 3D structure that could compensate for a weak input. Each frame's point cloud is sub-sampled to a fixed cardinality $P$ ($P{=}8192$ for Cheese3D, $P{=}1024$ for Chickadee), optionally centered and unit-normalized, and encoded by a per-frame PointNet~\citep{qi2017pointnet}: a shared MLP $(3\!\to\!64\!\to\!128\!\to\!256)$ with batch norm and ReLU is applied to every point, followed by global max pooling and a two-layer projection to a $d{=}256$ embedding. The resulting $(B,T,d)$ sequence of per-frame embeddings is processed by a 2-layer Transformer encoder (4 heads, GELU, pre-norm, learned positional embeddings up to length 256), and a linear readout maps each timestep to $N$ neuron-specific rate predictions. The output is passed through a softplus to enforce positivity since we train with a Poisson negative log-likelihood loss, $\mathcal{L} = \frac{1}{BTN}\sum (\hat{\lambda}_{btn} - y_{btn}\log\hat{\lambda}_{btn})$, where $y_{btn}$ is the observed spike count.

\paragraph{Training.}
We optimize with AdamW (weight decay $0.05$, gradient clip 1.0) for 300 epochs using a cosine schedule from the configured peak learning rate down to $10^{-6}$ (peak $10^{-5}$ for Cheese3D, $10^{-3}$ for Chickadee, set per dataset). The training batch size is 8 for the dense Cheese3D point clouds and 32 for the smaller Chickadee point clouds. We track validation bits-per-spike (BPS) every epoch and retain the checkpoint with the highest validation BPS for test-set evaluation. Because the train / val / test windows are non-overlapping in time and disjoint at the window level, a model that only memorized the training windows cannot trivially extrapolate to test windows.

\paragraph{Baseline.}
Holding the encoding architecture and training recipe fixed, we vary only the visual input so that any difference in test BPS reflects the input representation rather than the readout. Our model uses the foreground-filtered \beastthree point cloud, which is produced fully self-supervised: \beastthree is pretrained without any keypoint or pose annotations, and the per-frame point cloud is read out directly from the pretrained encoder without any task-specific labels. As a baseline, we replace this input with multi-view-triangulated 3D keypoints (outputs of the ensemble Kalman smoother (EKS)~\citep{biderman2024lightning}; 28 keypoints for Cheese3D, 18 for Chickadee; see Appendix~\ref{app:datasets}), which are produced by a supervised pose-estimation model trained on manually labeled keypoints in each view and then triangulated across views. This baseline therefore represents the prevailing label-dependent input modality for downstream behavioral analyses, whereas \beastthree requires no manual labeling.

\paragraph{Appearance control.}
The point clouds described above carry geometry only. Because every retained Gaussian is pixel-aligned, each 3D point can additionally be assigned the foreground-masked RGB color of the image pixel from which it was predicted, which makes it possible to separate the contribution of geometry from that of appearance. Holding the point set, the encoding model, and the training procedure fixed, we compare three per-point input representations: the 3D coordinates alone (XYZ, $P{\times}3$), the color alone (RGB, $P{\times}3$), and their concatenation (XYZ\,+\,RGB, $P{\times}6$). The RGB-only model has no access to spatial structure and thus serves as an appearance-only control. Table~\ref{tab:ephys_appearance} reports test BPS for both datasets. Color alone falls below the mean-rate null model on both datasets, and appending color to the coordinates leaves performance essentially unchanged. The encoding performance of \beastthree is therefore attributable to the recovered 3D structure rather than to image appearance.

\begin{table}[h]
\centering
\small
\caption{Appearance control for neural encoding. Test bits-per-spike (BPS, averaged over units) of the encoding model with the 3D coordinates (XYZ), the per-point color (RGB), or both (XYZ\,+\,RGB) as input. $P$ denotes the number of points per frame ($P{=}8192$ for Cheese3D, $P{=}1024$ for Chickadee). Learning rates are selected on the validation split.}
\label{tab:ephys_appearance}
\begin{tabular}{llr}
\toprule
\textbf{Model input} & \textbf{Input dim} & \textbf{Test BPS} \\
\midrule
\multicolumn{3}{l}{\textit{Cheese3D}} \\
XYZ & $P{\times}3$ & $0.3903$ \\
RGB & $P{\times}3$ & $-0.0430$ \\
XYZ\,+\,RGB & $P{\times}6$ & $0.3903$ \\
\midrule
\multicolumn{3}{l}{\textit{Chickadee}} \\
XYZ & $P{\times}3$ & $0.1141$ \\
RGB & $P{\times}3$ & $-0.0090$ \\
XYZ\,+\,RGB & $P{\times}6$ & $0.1188$ \\
\bottomrule
\end{tabular}
\end{table}

\section{Broader impacts} \label{app:impacts}
The \beastthree framework enables more efficient extraction of meaningful information from multi-view video data, potentially accelerating behavioral neuroscience research with several beneficial outcomes. By reducing the need for extensive human labeling while improving accuracy, \beastthree can democratize advanced video analysis capabilities for laboratories with limited resources. 
This efficiency could accelerate basic science discoveries that underlie advances in biomedical applications, neurological disorder treatments, and improved understanding of brain function.
While \beastthree is developed primarily for behavioral neuroscience studies using animal subjects, the
underlying technology could potentially be repurposed for human video analysis, raising several concerns:

\begin{itemize}
    \item Surveillance capabilities: The improved ability to track behaviors could enhance surveillance technologies, potentially infringing on privacy rights if deployed without appropriate oversight.
    \item Bias and fairness: As with any AI system trained on specific datasets, \beastthree-derived models may perform differently across demographic groups if applied to human subjects, potentially perpetuating biases in downstream applications.
    \item Resource inequality: While a pretrained \beastthree model can improve the efficiency of downstream tasks, the computational requirements for pretraining itself may limit access to this technology for under-resourced institutions, potentially widening existing disparities in research capabilities.
\end{itemize}

%% file: appendix/nvs_tables.tex

\begin{table}[t]
\centering
\small
\caption{Novel view synthesis on the In-Distribution Test (InD Test) set (foreground masked with the GT mask). \beastthree~(joint) is a single model trained on Cheese3D, Rat7M and Human3.6M (Appendix~\ref{app:nvs:joint}) and is not evaluated on Chickadee. $\uparrow$ / $\downarrow$: higher / lower is better.}
\label{tab:nvs_ind_gt}
\begin{tabular}{llccc}
\toprule
Dataset & Model & PSNR$\uparrow$ & SSIM$\uparrow$ & LPIPS$\downarrow$ \\
\midrule
\multirow{7}{*}{Cheese3D} & E-RayZer (ZS) & $13.645 \pm 2.834$ & $0.575 \pm 0.110$ & $0.467 \pm 0.051$ \\
 & E-RayZer (FT) & $13.131 \pm 2.876$ & $0.565 \pm 0.098$ & $0.473 \pm 0.057$ \\
 & Pose Splatter & $12.443 \pm 3.404$ & $0.594 \pm 0.113$ & $0.419 \pm 0.054$ \\
 & \beastthree~(no frustum) & $26.578 \pm 2.698$ & $0.837 \pm 0.043$ & $0.224 \pm 0.038$ \\
 & \beastthree~(no DINOv3) & $26.388 \pm 2.457$ & $0.832 \pm 0.044$ & $0.208 \pm 0.031$ \\
 & \beastthree~(joint) & $\mathbf{27.040} \pm 3.031$ & $\mathbf{0.855} \pm 0.041$ & $\mathbf{0.177} \pm 0.034$ \\
 & \beastthree & $26.990 \pm 2.659$ & $0.846 \pm 0.042$ & $0.194 \pm 0.032$ \\
\midrule
\multirow{7}{*}{Rat7M} & E-RayZer (ZS) & $12.536 \pm 3.271$ & $0.772 \pm 0.075$ & $0.235 \pm 0.048$ \\
 & E-RayZer (FT) & $12.398 \pm 3.314$ & $0.764 \pm 0.081$ & $0.237 \pm 0.049$ \\
 & Pose Splatter & $4.519 \pm 1.845$ & $0.673 \pm 0.092$ & $0.223 \pm 0.057$ \\
 & \beastthree~(no frustum) & $22.005 \pm 1.934$ & $0.900 \pm 0.038$ & $0.113 \pm 0.032$ \\
 & \beastthree~(no DINOv3) & $22.260 \pm 1.947$ & $0.906 \pm 0.038$ & $0.105 \pm 0.030$ \\
 & \beastthree~(joint) & $\mathbf{22.468} \pm 1.950$ & $\mathbf{0.914} \pm 0.038$ & $\mathbf{0.092} \pm 0.029$ \\
 & \beastthree & $22.423 \pm 1.902$ & $0.909 \pm 0.036$ & $0.102 \pm 0.029$ \\
\midrule
\multirow{6}{*}{Chickadee} & E-RayZer (ZS) & $10.173 \pm 3.041$ & $0.631 \pm 0.102$ & $0.298 \pm 0.060$ \\
 & E-RayZer (FT) & $9.537 \pm 3.062$ & $0.628 \pm 0.098$ & $0.298 \pm 0.058$ \\
 & Pose Splatter & $9.064 \pm 3.172$ & $0.603 \pm 0.095$ & $0.299 \pm 0.059$ \\
 & \beastthree~(no frustum) & $16.760 \pm 1.273$ & $0.708 \pm 0.070$ & $0.239 \pm 0.043$ \\
 & \beastthree~(no DINOv3) & $18.293 \pm 1.783$ & $0.767 \pm 0.069$ & $0.187 \pm 0.037$ \\
 & \beastthree & $\mathbf{18.592} \pm 1.751$ & $\mathbf{0.771} \pm 0.064$ & $\mathbf{0.186} \pm 0.038$ \\
\midrule
\multirow{7}{*}{Human3.6M} & E-RayZer (ZS) & $11.893 \pm 2.591$ & $0.769 \pm 0.061$ & $0.259 \pm 0.054$ \\
 & E-RayZer (FT) & $13.231 \pm 1.733$ & $0.786 \pm 0.053$ & $0.258 \pm 0.056$ \\
 & Pose Splatter & $9.648 \pm 2.335$ & $0.743 \pm 0.063$ & $0.191 \pm 0.048$ \\
 & \beastthree~(no frustum) & $22.206 \pm 2.080$ & $0.929 \pm 0.031$ & $0.098 \pm 0.036$ \\
 & \beastthree~(no DINOv3) & $\mathbf{22.701} \pm 2.007$ & $\mathbf{0.934} \pm 0.029$ & $0.091 \pm 0.035$ \\
 & \beastthree~(joint) & $22.450 \pm 2.380$ & $\mathbf{0.934} \pm 0.032$ & $\mathbf{0.086} \pm 0.034$ \\
 & \beastthree & $22.386 \pm 2.169$ & $0.931 \pm 0.031$ & $0.092 \pm 0.035$ \\
\bottomrule
\end{tabular}
\end{table}

\begin{table}[t]
\centering
\small
\caption{Novel view synthesis on the Out-of-Distribution Test (OOD Test) set (foreground masked with the GT mask). \beastthree~(joint) is a single model trained on Cheese3D, Rat7M and Human3.6M (Appendix~\ref{app:nvs:joint}) and is not evaluated on Chickadee.}
\label{tab:nvs_ood_gt}
\begin{tabular}{llccc}
\toprule
Dataset & Model & PSNR$\uparrow$ & SSIM$\uparrow$ & LPIPS$\downarrow$ \\
\midrule
\multirow{7}{*}{Cheese3D} & E-RayZer (ZS) & $12.981 \pm 2.848$ & $0.578 \pm 0.101$ & $0.467 \pm 0.041$ \\
 & E-RayZer (FT) & $12.307 \pm 2.752$ & $0.563 \pm 0.082$ & $0.480 \pm 0.047$ \\
 & Pose Splatter & $12.452 \pm 2.873$ & $0.592 \pm 0.095$ & $0.420 \pm 0.043$ \\
 & \beastthree~(no frustum) & $20.175 \pm 1.852$ & $0.712 \pm 0.050$ & $0.298 \pm 0.033$ \\
 & \beastthree~(no DINOv3) & $\mathbf{21.220} \pm 2.060$ & $\mathbf{0.732} \pm 0.049$ & $0.281 \pm 0.025$ \\
 & \beastthree~(joint) & $20.476 \pm 2.059$ & $0.728 \pm 0.051$ & $\mathbf{0.257} \pm 0.028$ \\
 & \beastthree & $20.736 \pm 2.000$ & $0.720 \pm 0.047$ & $0.273 \pm 0.030$ \\
\midrule
\multirow{7}{*}{Rat7M} & E-RayZer (ZS) & $12.322 \pm 2.650$ & $0.767 \pm 0.078$ & $0.235 \pm 0.048$ \\
 & E-RayZer (FT) & $12.889 \pm 2.682$ & $0.751 \pm 0.087$ & $0.259 \pm 0.066$ \\
 & Pose Splatter & $5.188 \pm 2.200$ & $0.682 \pm 0.094$ & $0.221 \pm 0.058$ \\
 & \beastthree~(no frustum) & $\mathbf{16.265} \pm 2.569$ & $\mathbf{0.811} \pm 0.061$ & $0.153 \pm 0.043$ \\
 & \beastthree~(no DINOv3) & $15.936 \pm 2.510$ & $0.810 \pm 0.060$ & $0.151 \pm 0.042$ \\
 & \beastthree~(joint) & $13.883 \pm 2.574$ & $0.752 \pm 0.065$ & $0.186 \pm 0.046$ \\
 & \beastthree & $16.006 \pm 2.360$ & $0.804 \pm 0.061$ & $\mathbf{0.147} \pm 0.044$ \\
\midrule
\multirow{6}{*}{Chickadee} & E-RayZer (ZS) & $10.500 \pm 3.110$ & $\mathbf{0.667} \pm 0.105$ & $0.283 \pm 0.063$ \\
 & E-RayZer (FT) & $9.454 \pm 2.959$ & $0.664 \pm 0.104$ & $0.287 \pm 0.066$ \\
 & Pose Splatter & $8.601 \pm 1.868$ & $0.651 \pm 0.090$ & $0.253 \pm 0.052$ \\
 & \beastthree~(no frustum) & $8.664 \pm 1.618$ & $0.523 \pm 0.125$ & $0.329 \pm 0.080$ \\
 & \beastthree~(no DINOv3) & $9.771 \pm 1.668$ & $0.547 \pm 0.109$ & $0.297 \pm 0.065$ \\
 & \beastthree & $\mathbf{13.726} \pm 1.746$ & $0.650 \pm 0.105$ & $\mathbf{0.235} \pm 0.050$ \\
\midrule
\multirow{7}{*}{Human3.6M} & E-RayZer (ZS) & $11.613 \pm 3.623$ & $0.724 \pm 0.083$ & $0.274 \pm 0.056$ \\
 & E-RayZer (FT) & $12.795 \pm 2.495$ & $0.749 \pm 0.064$ & $0.275 \pm 0.062$ \\
 & Pose Splatter & $8.560 \pm 2.472$ & $0.698 \pm 0.068$ & $0.225 \pm 0.047$ \\
 & \beastthree~(no frustum) & $17.214 \pm 1.662$ & $0.806 \pm 0.051$ & $0.179 \pm 0.037$ \\
 & \beastthree~(no DINOv3) & $\mathbf{18.712} \pm 1.984$ & $\mathbf{0.830} \pm 0.052$ & $\mathbf{0.159} \pm 0.039$ \\
 & \beastthree~(joint) & $16.447 \pm 1.687$ & $0.784 \pm 0.056$ & $0.192 \pm 0.038$ \\
 & \beastthree & $17.107 \pm 1.697$ & $0.801 \pm 0.052$ & $0.180 \pm 0.037$ \\
\bottomrule
\end{tabular}
\end{table}

\begin{table}[t]
\centering
\small
\caption{Novel view synthesis and foreground segmentation on the In-Distribution Test (InD Test) set (rendered alpha as the foreground mask).}
\label{tab:nvs_ind_pred}
\resizebox{\textwidth}{!}{%
\begin{tabular}{llcccc}
\toprule
Dataset & Model & PSNR$\uparrow$ & SSIM$\uparrow$ & LPIPS$\downarrow$ & IoU$\uparrow$ \\
\midrule
\multirow{4}{*}{Cheese3D} & Pose Splatter & $15.255 \pm 2.602$ & $0.580 \pm 0.109$ & $0.440 \pm 0.052$ & $0.740 \pm 0.079$ \\
 & \beastthree~(no frustum) & $26.087 \pm 2.856$ & $0.831 \pm 0.046$ & $0.228 \pm 0.039$ & $\mathbf{0.933} \pm 0.059$ \\
 & \beastthree~(no DINOv3) & $25.969 \pm 2.649$ & $0.827 \pm 0.047$ & $0.212 \pm 0.033$ & $0.899 \pm 0.058$ \\
 & \beastthree & $\mathbf{26.623} \pm 2.838$ & $\mathbf{0.842} \pm 0.044$ & $\mathbf{0.197} \pm 0.033$ & $0.903 \pm 0.058$ \\
\midrule
\multirow{4}{*}{Rat7M} & Pose Splatter & $9.430 \pm 2.127$ & $0.580 \pm 0.103$ & $0.377 \pm 0.064$ & $0.129 \pm 0.149$ \\
 & \beastthree~(no frustum) & $21.956 \pm 1.965$ & $0.900 \pm 0.039$ & $0.113 \pm 0.032$ & $0.500 \pm 0.087$ \\
 & \beastthree~(no DINOv3) & $22.160 \pm 1.994$ & $0.905 \pm 0.038$ & $0.106 \pm 0.030$ & $0.548 \pm 0.097$ \\
 & \beastthree & $\mathbf{22.327} \pm 1.945$ & $\mathbf{0.908} \pm 0.037$ & $\mathbf{0.103} \pm 0.030$ & $\mathbf{0.551} \pm 0.093$ \\
\midrule
\multirow{4}{*}{Chickadee} & Pose Splatter & $10.256 \pm 2.491$ & $0.499 \pm 0.134$ & $0.395 \pm 0.075$ & $0.259 \pm 0.164$ \\
 & \beastthree~(no frustum) & $16.640 \pm 1.366$ & $0.706 \pm 0.071$ & $0.242 \pm 0.043$ & $0.547 \pm 0.055$ \\
 & \beastthree~(no DINOv3) & $18.029 \pm 1.945$ & $0.764 \pm 0.072$ & $0.191 \pm 0.039$ & $\mathbf{0.692} \pm 0.085$ \\
 & \beastthree & $\mathbf{18.314} \pm 1.910$ & $\mathbf{0.768} \pm 0.066$ & $\mathbf{0.189} \pm 0.039$ & $0.681 \pm 0.081$ \\
\midrule
\multirow{4}{*}{Human3.6M} & Pose Splatter & $13.998 \pm 1.688$ & $0.731 \pm 0.068$ & $0.224 \pm 0.064$ & $0.581 \pm 0.120$ \\
 & \beastthree~(no frustum) & $22.102 \pm 2.126$ & $0.928 \pm 0.032$ & $0.098 \pm 0.036$ & $0.695 \pm 0.055$ \\
 & \beastthree~(no DINOv3) & $\mathbf{22.624} \pm 2.046$ & $\mathbf{0.933} \pm 0.029$ & $\mathbf{0.091} \pm 0.035$ & $\mathbf{0.718} \pm 0.056$ \\
 & \beastthree & $22.272 \pm 2.223$ & $0.930 \pm 0.032$ & $0.093 \pm 0.035$ & $0.710 \pm 0.054$ \\
\bottomrule
\end{tabular}
}
\end{table}

\begin{table}[t]
\centering
\small
\caption{Novel view synthesis and foreground segmentation on the Out-of-Distribution Test (OOD Test) set (rendered alpha as the foreground mask).}
\label{tab:nvs_ood_pred}
\resizebox{\textwidth}{!}{%
\begin{tabular}{llcccc}
\toprule
Dataset & Model & PSNR$\uparrow$ & SSIM$\uparrow$ & LPIPS$\downarrow$ & IoU$\uparrow$ \\
\midrule
\multirow{4}{*}{Cheese3D} & Pose Splatter & $14.300 \pm 2.710$ & $0.572 \pm 0.095$ & $0.447 \pm 0.044$ & $0.740 \pm 0.062$ \\
 & \beastthree~(no frustum) & $19.569 \pm 1.750$ & $0.700 \pm 0.050$ & $0.305 \pm 0.032$ & $\mathbf{0.886} \pm 0.039$ \\
 & \beastthree~(no DINOv3) & $\mathbf{20.607} \pm 2.068$ & $\mathbf{0.720} \pm 0.051$ & $0.292 \pm 0.026$ & $0.874 \pm 0.043$ \\
 & \beastthree & $20.209 \pm 1.879$ & $0.709 \pm 0.047$ & $\mathbf{0.280} \pm 0.031$ & $0.877 \pm 0.044$ \\
\midrule
\multirow{4}{*}{Rat7M} & Pose Splatter & $9.928 \pm 2.165$ & $0.607 \pm 0.100$ & $0.356 \pm 0.062$ & $0.175 \pm 0.136$ \\
 & \beastthree~(no frustum) & $\mathbf{15.853} \pm 2.716$ & $\mathbf{0.802} \pm 0.064$ & $0.170 \pm 0.049$ & $0.343 \pm 0.135$ \\
 & \beastthree~(no DINOv3) & $15.330 \pm 2.617$ & $0.798 \pm 0.062$ & $0.174 \pm 0.047$ & $0.355 \pm 0.157$ \\
 & \beastthree & $15.507 \pm 2.453$ & $0.792 \pm 0.065$ & $\mathbf{0.169} \pm 0.051$ & $\mathbf{0.400} \pm 0.112$ \\
\midrule
\multirow{4}{*}{Chickadee} & Pose Splatter & $9.943 \pm 1.747$ & $0.560 \pm 0.105$ & $0.355 \pm 0.062$ & $0.382 \pm 0.130$ \\
 & \beastthree~(no frustum) & $7.933 \pm 1.443$ & $0.494 \pm 0.129$ & $0.389 \pm 0.081$ & $0.105 \pm 0.080$ \\
 & \beastthree~(no DINOv3) & $8.779 \pm 1.460$ & $0.514 \pm 0.115$ & $0.359 \pm 0.067$ & $0.148 \pm 0.074$ \\
 & \beastthree & $\mathbf{13.016} \pm 1.817$ & $\mathbf{0.634} \pm 0.107$ & $\mathbf{0.258} \pm 0.054$ & $\mathbf{0.462} \pm 0.113$ \\
\midrule
\multirow{4}{*}{Human3.6M} & Pose Splatter & $15.281 \pm 1.569$ & $0.685 \pm 0.073$ & $0.265 \pm 0.064$ & $0.562 \pm 0.151$ \\
 & \beastthree~(no frustum) & $16.888 \pm 1.731$ & $0.794 \pm 0.056$ & $0.185 \pm 0.040$ & $0.635 \pm 0.067$ \\
 & \beastthree~(no DINOv3) & $\mathbf{18.427} \pm 2.018$ & $\mathbf{0.822} \pm 0.055$ & $\mathbf{0.163} \pm 0.041$ & $0.625 \pm 0.068$ \\
 & \beastthree & $16.741 \pm 1.764$ & $0.788 \pm 0.057$ & $0.187 \pm 0.040$ & $\mathbf{0.640} \pm 0.068$ \\
\bottomrule
\end{tabular}
}
\end{table}

%% file: checklist.tex
\section*{NeurIPS Paper Checklist}

\begin{enumerate}

\item {\bf Claims}
    \item[] Question: Do the main claims made in the abstract and introduction accurately reflect the paper's contributions and scope?
    \item[] Answer: \answerYes{}
    \item[] Justification: We clearly declare and introduce our main contribution as a self-supervised 3D Gaussian splat technique for animal behavior videos.

\item {\bf Limitations}
    \item[] Question: Does the paper discuss the limitations of the work performed by the authors?
    \item[] Answer: \answerYes{}
    \item[] Justification: See Conclusion section for limitations of the work.

\item {\bf Theory assumptions and proofs}
    \item[] Question: For each theoretical result, does the paper provide the full set of assumptions and a complete (and correct) proof?
    \item[] Answer: \answerNA{}
    \item[] Justification: This paper does not include theoretical results.

    \item {\bf Experimental result reproducibility}
    \item[] Question: Does the paper fully disclose all the information needed to reproduce the main experimental results of the paper to the extent that it affects the main claims and/or conclusions of the paper (regardless of whether the code and data are provided or not)?
    \item[] Answer: \answerYes{}
    \item[] Justification: We include model architecture and training details in the main text and Appendix. We also provide hyperparameter ranges and selection strategy for all model training experiments.

\item {\bf Open access to data and code}
    \item[] Question: Does the paper provide open access to the data and code, with sufficient instructions to faithfully reproduce the main experimental results, as described in supplemental material?
    \item[] Answer: \answerYes{}
    \item[] Justification: Code and data preparation instructions are available at \url{https://github.com/paninski-lab/beast}. URLs for public datasets are in the Appendix.

\item {\bf Experimental setting/details}
    \item[] Question: Does the paper specify all the training and test details (e.g., data splits, hyperparameters, how they were chosen, type of optimizer) necessary to understand the results?
    \item[] Answer: \answerYes{}
    \item[] Justification: We provide all training and hyperparameter details in the Appendix.

\item {\bf Experiment statistical significance}
    \item[] Question: Does the paper report error bars suitably and correctly defined or other appropriate information about the statistical significance of the experiments?
    \item[] Answer: \answerYes{}
    \item[] Justification: We report error bars in tables and plots and describe how they are calculated in the corresponding captions or accompanying text.

\item {\bf Experiments compute resources}
    \item[] Question: For each experiment, does the paper provide sufficient information on the computer resources (type of compute workers, memory, time of execution) needed to reproduce the experiments?
    \item[] Answer: \answerNo{}
    \item[] Justification: We report an estimate of the required compute resources for initial pretraining in the Conclusion, but do not provide compute resource estimates for the downstream tasks.
    
\item {\bf Code of ethics}
    \item[] Question: Does the research conducted in the paper conform, in every respect, with the NeurIPS Code of Ethics \url{https://neurips.cc/public/EthicsGuidelines}?
    \item[] Answer: \answerYes{}
    \item[] Justification: We have carefully read and followed the NeurIPS Code of Ethics.

\item {\bf Broader impacts}
    \item[] Question: Does the paper discuss both potential positive societal impacts and negative societal impacts of the work performed?
    \item[] Answer: \answerYes{}
    \item[] Justification: We discuss broader impacts in the Appendix.
    
\item {\bf Safeguards}
    \item[] Question: Does the paper describe safeguards that have been put in place for responsible release of data or models that have a high risk for misuse (e.g., pre-trained language models, image generators, or scraped datasets)?
    \item[] Answer: \answerNA{}
    \item[] Justification: The paper does not include any data or models that have a high risk for misuse.

\item {\bf Licenses for existing assets}
    \item[] Question: Are the creators or original owners of assets (e.g., code, data, models), used in the paper, properly credited and are the license and terms of use explicitly mentioned and properly respected?
    \item[] Answer: \answerYes{}
    \item[] Justification: For all existing assets we clearly cite the original paper and provide the name of the license.

\item {\bf New assets}
    \item[] Question: Are new assets introduced in the paper well documented and is the documentation provided alongside the assets?
    \item[] Answer: \answerNA{}
    \item[] Justification: This paper does not release new assets.

\item {\bf Crowdsourcing and research with human subjects}
    \item[] Question: For crowdsourcing experiments and research with human subjects, does the paper include the full text of instructions given to participants and screenshots, if applicable, as well as details about compensation (if any)? 
    \item[] Answer: \answerNA{}
    \item[] Justification: This paper does not involve crowdsourcing or research with human subjects.

\item {\bf Institutional review board (IRB) approvals or equivalent for research with human subjects}
    \item[] Question: Does the paper describe potential risks incurred by study participants, whether such risks were disclosed to the subjects, and whether Institutional Review Board (IRB) approvals (or an equivalent approval/review based on the requirements of your country or institution) were obtained?
    \item[] Answer: \answerNA{}
    \item[] Justification: This paper does not involve research with human subjects.

\item {\bf Declaration of LLM usage}
    \item[] Question: Does the paper describe the usage of LLMs if it is an important, original, or non-standard component of the core methods in this research? Note that if the LLM is used only for writing, editing, or formatting purposes and does \emph{not} impact the core methodology, scientific rigor, or originality of the research, declaration is not required.
    \item[] Answer: \answerNA{}
    \item[] Justification: The core method development in this paper does not involve LLMs.

\end{enumerate}